\documentclass{rspublic}
\usepackage{graphicx}
\usepackage{amssymb,amsmath}
\usepackage{natbib}

\begin{document}

\title[Ternary probabilistic forecasts]{On the visualisation, verification and recalibration of ternary probabilistic forecasts}
\author[T.E. Jupp et al.]{Tim E. Jupp$^{1}$\footnote{Author for correspondence (t.e.jupp@ex.ac.uk)}, Rachel Lowe$^{2}$, \\                                                            Caio A.S. Coelho$^3$, David B. Stephenson$^1$}
\affiliation{$^1$Mathematics Research Institute, University of Exeter, Exeter, UK. \\
$^2$Abdus Salam International Centre for Theoretical Physics, Trieste, Italy. \\
$^3$Centro de Previs\~ao de Tempo e Estudos Clim\'aticos,\\
Instituto Nacional de Pesquisas Espaciais,\\
Cachoeira Paulista, 12630-000, SP, Brazil.
}
\label{firstpage}
\maketitle

\begin{abstract}{map climate colour}

We develop a graphical interpretation of ternary probabilistic forecasts in which forecasts and observations are regarded as points inside a triangle. Within the triangle, we define a continuous colour palette in which hue and colour saturation are defined with reference to the observed climatology. In contrast to current methods, forecast maps created with this colour scheme convey all of the information present in each ternary forecast.

The geometrical interpretation is then extended to verification under quadratic scoring rules (of which the Brier Score and the Ranked Probability Score are well--known examples). Each scoring rule defines an associated triangle in which the square roots of the \emph{score}, the \emph{reliability}, the \emph{uncertainty} and the \emph{resolution} all have natural interpretations as root--mean--square distances. This leads to our proposal for a \emph{Ternary Reliability Diagram} in which data relating to verification and calibration can be summarised.

We illustrate these ideas with data relating to seasonal forecasting of precipitation in South America, including an example of nonlinear forecast calibration. Codes implementing these ideas have been produced using the statistical software package \texttt{R} and are available from the authors.
\end{abstract}

\section{Introduction}\label{sec:intro}

Forecasts are often given in probabilistic terms. For example: `\emph{There is a 60\% chance that rainfall before 12h00 tomorrow will exceed} $15\mathrm{mm}$' ; `\emph{The odds for the football match are 6--to--1 against a home win and even money for a draw}' ; `\emph{The climate model predicts that mean daytime temperature in summer 2100 will be normally distributed with mean $\mu$ and variance $\sigma^{2}$'}. In each of these examples the forecast consists of a probability distribution. The sample spaces for the observable events of interest consist of two elements (exceed/not exceed), three elements (home win/draw/away win) and the real line $\mathbb{R}$. Consequently, these examples relate to binary, ternary and continuous probabilistic forecasts respectively.

This paper is concerened with two types of visualisation. In physical space, visualisation constitutes colour maps of probabilistic forecasts over a geographical region. In probability space, visualisation will constitute a geometrical representation of forecasts and observations in our proposed Ternary Reliability Diagram.

For binary forecasts, visualisation tools such as the reliability diagram, the sharpness diagram and the relative operating characteristic (ROC) curve are well known \cite{Jolliffe03}. Our aim in this paper is to develop analogues of these ideas for ternary forecasts and to develop a geometrical intuition for verification and recalibration.

The structure of this paper is as follows. \S \ref{sec:prob_forecast} contains a discussion of probabilistic forecasting, and illustrates how a continuous forecast distribution can be identified with a ternary forecast. \S \ref{sec:barycentric} then introduces the idea of representing each ternary forecast as a point in barycentric coordinates, and hence assigning to it a unique colour. In \S \ref{sec:verify} the quality of a forecasting system is quantified using quadratic scoring rules. This is then used in \S \ref{sec:calib} to propose an algorithm for recalibrating probabilistic forecasts which is illustrated with seasonal forecasting data.  Conclusions are presented in \S \ref{sec:conclusions}.

\section{Probabilistic forecasting}\label{sec:prob_forecast}

The generic situation to be considered is as follows: a forecasting system (perhaps a suite of climate models with different initial conditions) can be used to produce probabilistic forecasts (that is, probability distributions) for some variable of interest at a number of points within a geographical area. The subsequent challenge is to display in map form as much as possible of the information contained in this spatial array of probability distributions. The most natural thing to do, perhaps, is to associate with each forecast distribution a set of scalars (such as the distribution's moments or a selection of its quantiles) and produce a separate map for each scalar. The disadvantage of this procedure is that multiple maps are needed to convey the information contained in the forecast distributions. If information on the skill of the forecasting system were available, yet another map would be required.

\begin{figure}
\begin{center}
\includegraphics*[width=6cm,angle=0]{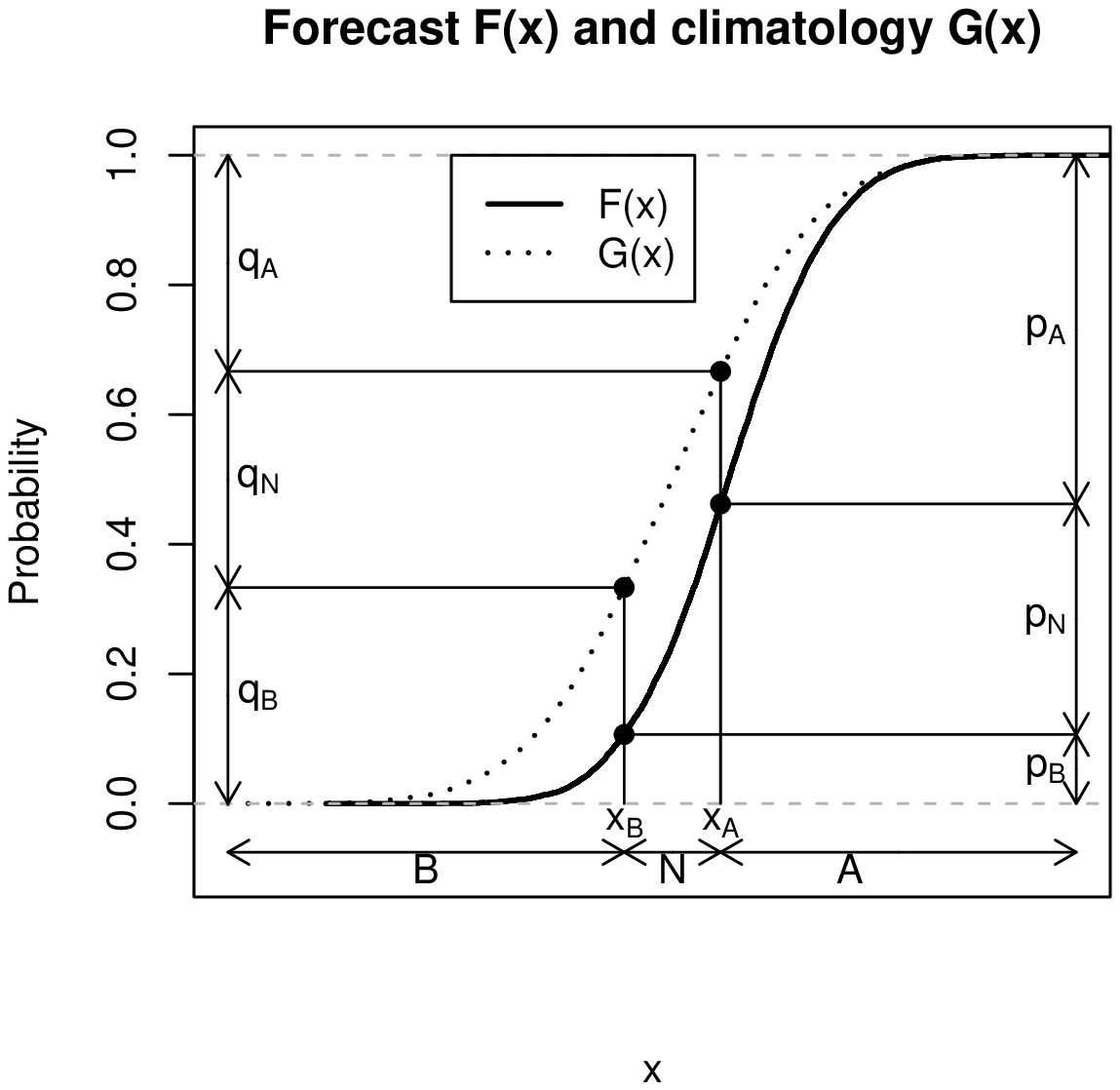}
\includegraphics*[width=6cm,angle=0]{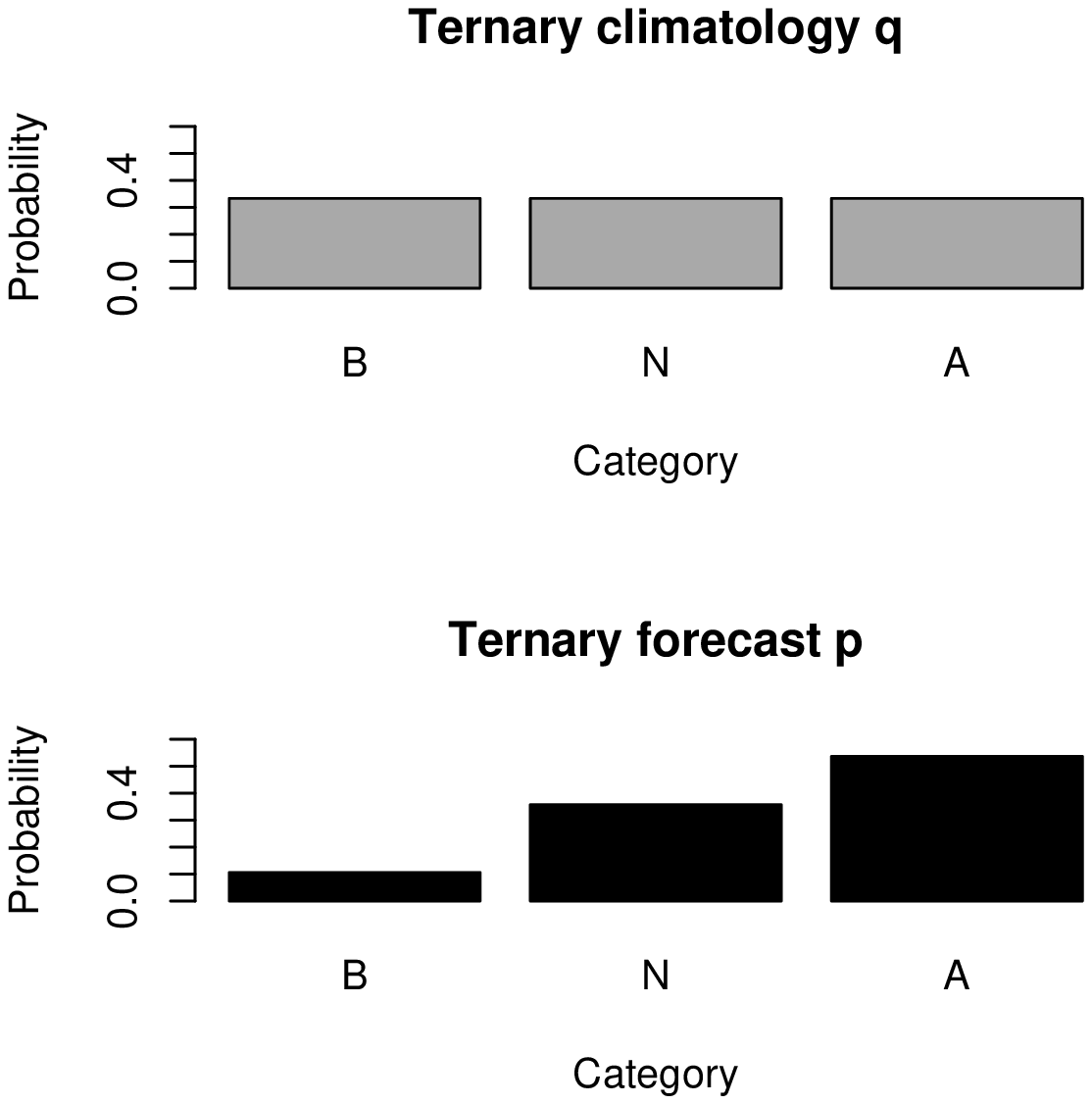}
\caption{Continuous distributions functions $F(x)$ (forecast) and $G(x)$ (climatology) can be associated with ternary forecasts $\mathbf{p}$ and $\mathbf{q}$ by defining three categories $B$, $N$ and $A$ -- `below--', `near--' and `above--' normal (Equations \ref{eq:q} and \ref{eq:t12}).}\label{fig:t}
\end{center}
\end{figure}

Consider probabilistic forecasting of a scalar climate variable $x$ such as temperature or precipitation. The forecasting system produces a cumulative distribution function (CDF) $F(x)$ at each spatial location (Figure \ref{fig:t}). The CDF expresses the forecast probability that the variable of interest will not exceed $x$. The forecast distribution $F(x)$ can be interpreted as the forecasting system's state of knowledge about the future value of the variable $x$. In contrast to a deterministic forecast $x_{f}$ (which is a scalar), a probabilistic forecast $F(x)$ is (in principle at least) an analytic function and hence an infinite--dimensional object. For this reason, it is not immediately obvious how to visualise spatial probabilistic forecasts.

Often, the variable $x$ has associated with it an observed climatology represented by the CDF $G(x)$ (Figure \ref{fig:t}). The climatology can be interpreted as the historical distribution of the observable quantity. Typically a climatology is calculated by aggregating data from an observational record over at least the past 30 years.

It is important to note that the climatology $G(x)$ and the forecast $F(x)$ are both continuous probability distributions and so it is quite possible for the forecasting system to issue the climatology $G(x)$ as its forecast. This is a perfectly valid probabilistic forecast but not, perhaps, a very useful one. It is usually hoped that the forecasting system's state of knowledge about the future extends beyond knowledge of the climatology alone. For this reason, it will be helpful to view the climatology as a benchmark distribution with which all other forecasts should be compared. Thus, in the discussion below, a colour will be assigned to the forecast $F(x)$ by considering the `distance' between the forecast $F(x)$ and the climatology $G(x)$.

\subsection{Ternary probabilistic forecasts}

In order to reduce the dimensionality of the problem it is useful to project the continuous distributions $G(x)$ and $F(x)$ onto ternary distributions $\mathbf{q}' = (q_{B},q_{N},q_{A})$ and $\mathbf{p}' = (p_{B},p_{N},p_{A})$ (Figure \ref{fig:t}).  We adopt the convention that vectors are column vectors and that $'$ denotes a transpose. The real line is divided into three ordered categories $B = (-\infty,x_{B}]$, $N = [x_{B},x_{A}]$ and $A = [x_{A},\infty)$ whose labels have the following rationale: $B$ -- `below normal', $N$ -- `near normal' and $A$ -- `above normal'.

The ternary distributions $\mathbf{q}(G)$ and $\mathbf{p}(F)$ encode the probabilities with which the variable is forecast to lie in the categories $B$, $N$ and $A$ (Figure \ref{fig:t}), and so all of their elements are non--negative with $q_{B}+q_{N}+q_{A}=p_{B}+p_{N}+p_{A}=1$.

For given quantiles $x_{B}$ and $x_{A}$ it follows that the ternary climatology $\mathbf{q}$ and ternary forecast $\mathbf{p}$ are given by:
\begin{equation}\label{eq:q}
\begin{array}{lllll}
\mathbf{q}'&=&(q_{B},q_{N},q_{A})& = &(G(x_{B}),G(x_{A})-G(x_{B}),1-G(x_{A})) \\
\mathbf{p}'&=&(p_{B},p_{N},p_{A})& = &(F(x_{B}),F(x_{A})-F(x_{B}),1-F(x_{A}))
\end{array}
\end{equation}

An equivalent interpretation is that for a specified ternary climatology $\mathbf{q}$ the categories $B$, $N$ and $A$ are defined by the quantiles:
\begin{equation}\label{eq:t12}
x_{B}=G^{-1}(q_{B}); \quad x_{A}=G^{-1}(q_{B}+q_{N})
\end{equation}
A natural choice for the ternary climatology is the uniform distribution $\mathbf{q}'=(\frac{1}{3},\frac{1}{3},\frac{1}{3})$. In this case -- which we shall regard as the default -- the categories $B$, $N$ and $A$ are defined by the terciles of the climatology $G(x)$ and observations over a long period would be expected to lie with equal frequency in each of the three categories. It should be stressed, however, that any choice of quantiles can be made when defining the ternary climatology. For example, the choice $\mathbf{q}'=(\frac{1}{4},\frac{1}{2},\frac{1}{4})$ would imply that categories $B$ and $A$ had been chosen to be the lower and upper quartiles of the climatology.

\section{Barycentric coordinates}\label{sec:barycentric}

In this section, a geometrical representation of a ternary forecast $\mathbf{p}$, ternary climatology $\mathbf{q}$ and ternary observation $\mathbf{o}$ is considered.

\subsection{Scoring functions}\label{sec:scores}

Scoring functions quantify the past skill of a forecasting system. The idea is to compare a ternary forecast $\mathbf{p}$ with the corresponding ternary observation $\mathbf{o}$ made after the event. Since the categories are exclusive and complete, it follows that a ternary observation $\mathbf{o}$ can take one of three values $\mathbf{o}_{B}'= (1,0,0)$, $\mathbf{o}_{N}' = (0,1,0)$ and $\mathbf{o}_{A}' = (0,0,1)$ according to whether the observable is found to lie in category $B$, $N$ or $A$.

The skill of a forecasting system can be quantified for a set of forecast--observation pairs by a score function $S$. The score is a measure of the difference between forecasts and observations. It follows that the lower the score, the more skilful the forecast.

In this paper, attention is restricted to scores defined by quadratic forms
\begin{equation}\label{eq:score}
S = \overline{(\mathbf{p}-\mathbf{o})' L' L (\mathbf{p}-\mathbf{o})}
\end{equation}
where the 3--by--3 matrix $L$ defines the particular scoring rule being used (Appendix, section \ref{sec:brier}), $L'L$ is assumed to be positive definite and the overbar denotes an average over all forecast--observation pairs. The Brier Score (Appendix, section \ref{sec:brier}) and the Ranked Probability Score (Appendix, section \ref{sec:rps}) are well--known examples of quadratic scores \cite{Brier50, Epstein69, Murphy69, Murphy75, Staelvonholstein78}.

The scoring matrix $L$ can be used to define a 2--by--3 matrix $\hat{M}$ (Appendix, equation \ref{eq:M}) which maps a ternary forecast $\mathbf{p} \in \mathbb{R}^3$ to a corresponding point $\mathbf{P} = \hat{M}\mathbf{p} \in \mathbb{R}^2 $ in the plane. This transformation maps ternary forecasts to points within a triangle and so each scoring matrix $L$ induces a corresponding triangle in $\mathbb{R}^2 $. In this paper, the matrix $\hat{M}$ is defined so that a score $S$ corresponds to a (mean) squared distance between forecasts $\mathbf{P}$ and observations $\mathbf{O}$ (considered as points inside an appropriate triangle):
\begin{equation}\label{eq:scorein R2}
S = \overline{ \| \mathbf{P}-\mathbf{O} \|^{2} }
\end{equation}

\begin{figure}
\begin{center}
\includegraphics*[angle=0,width=6cm]{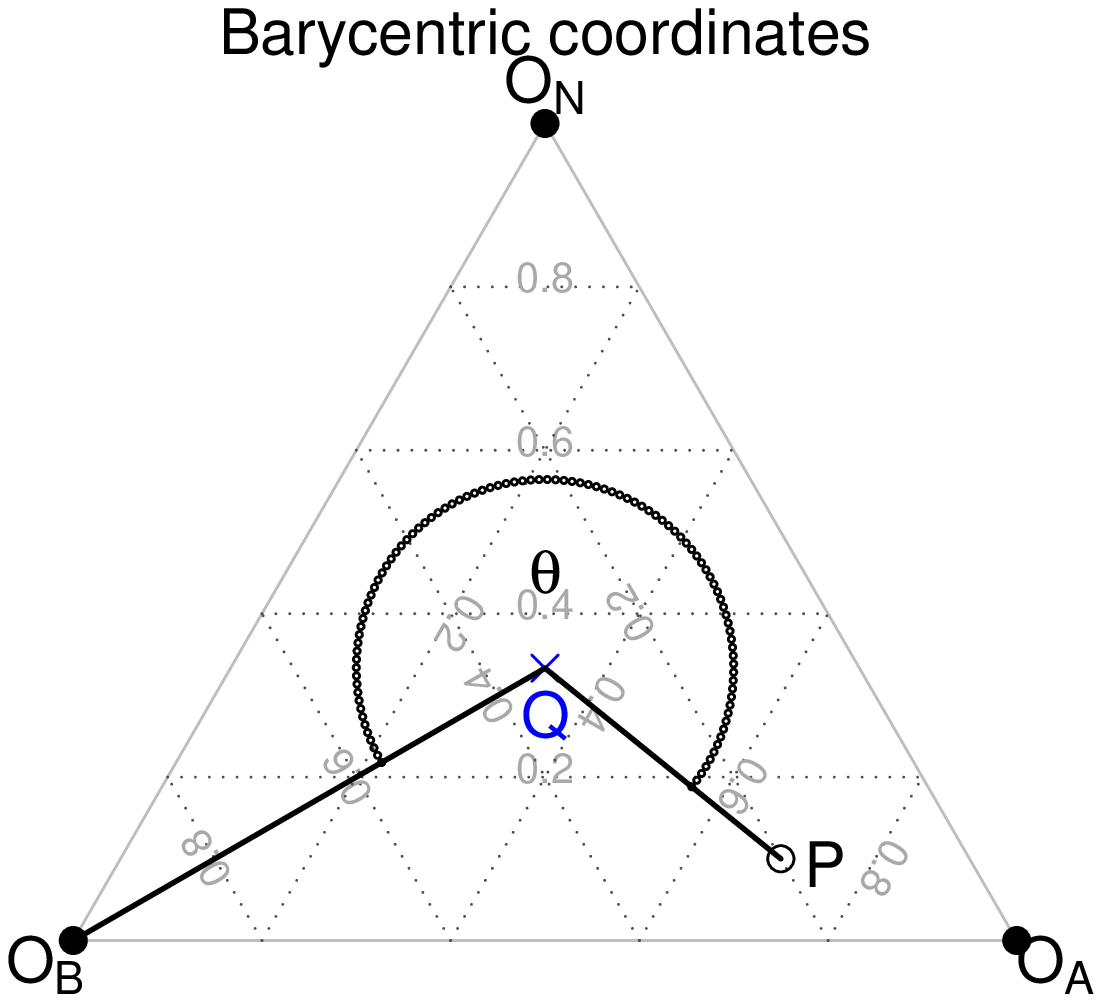}
\includegraphics*[angle=0,width=6cm]{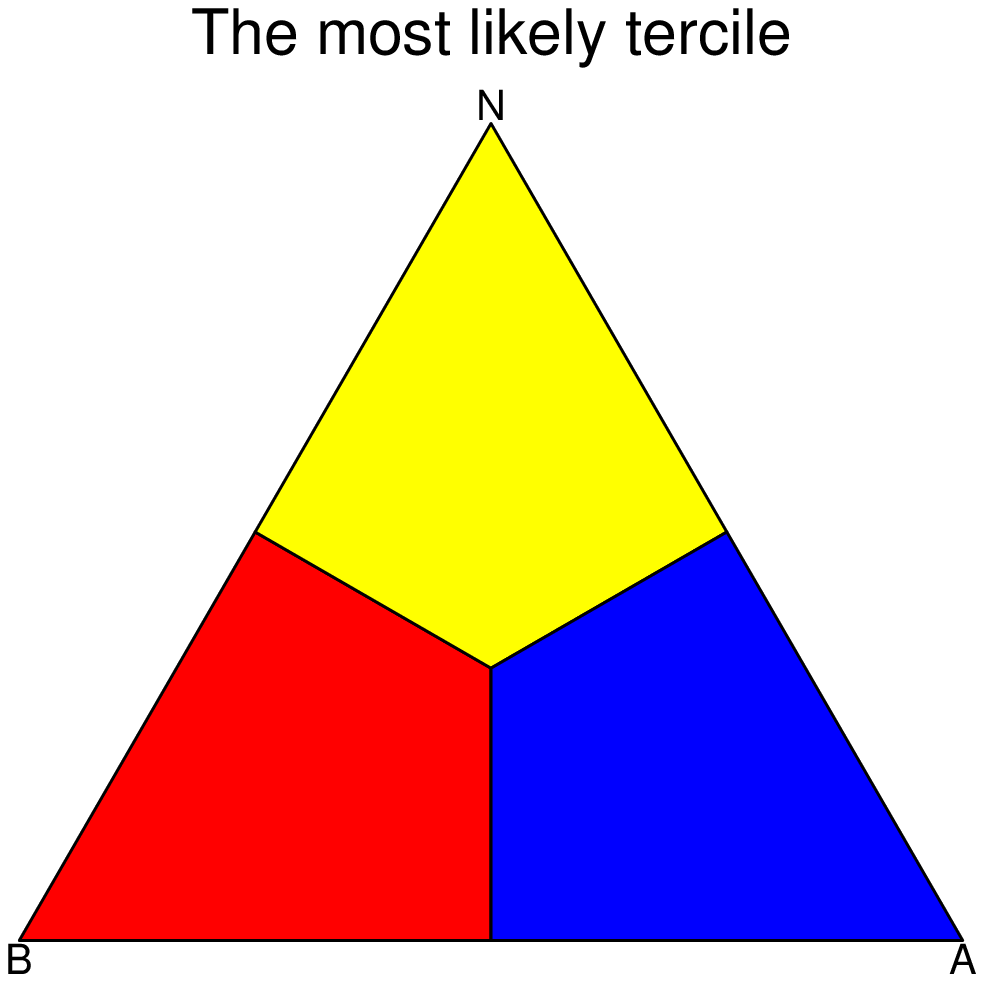}
\includegraphics*[angle=0,width=6cm]{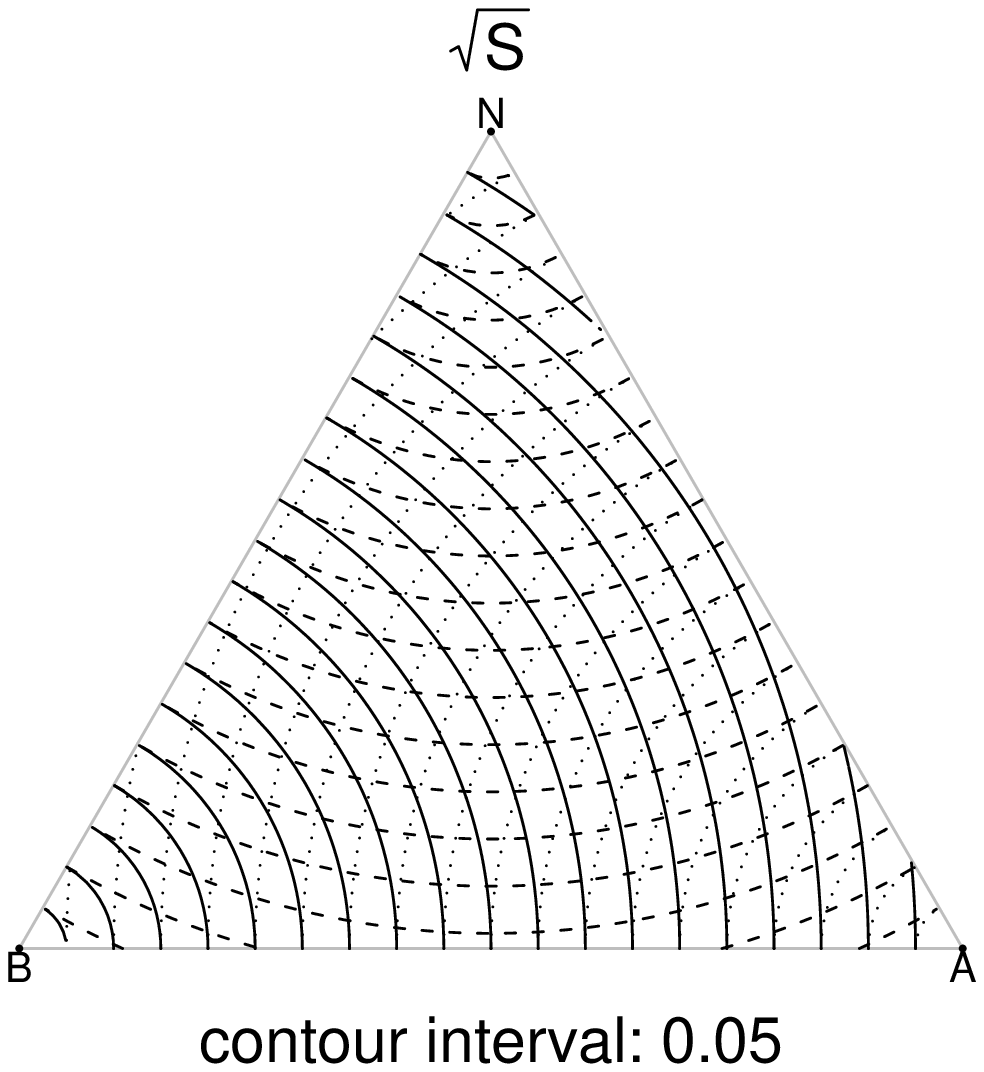}
\includegraphics*[angle=0,width=6cm]{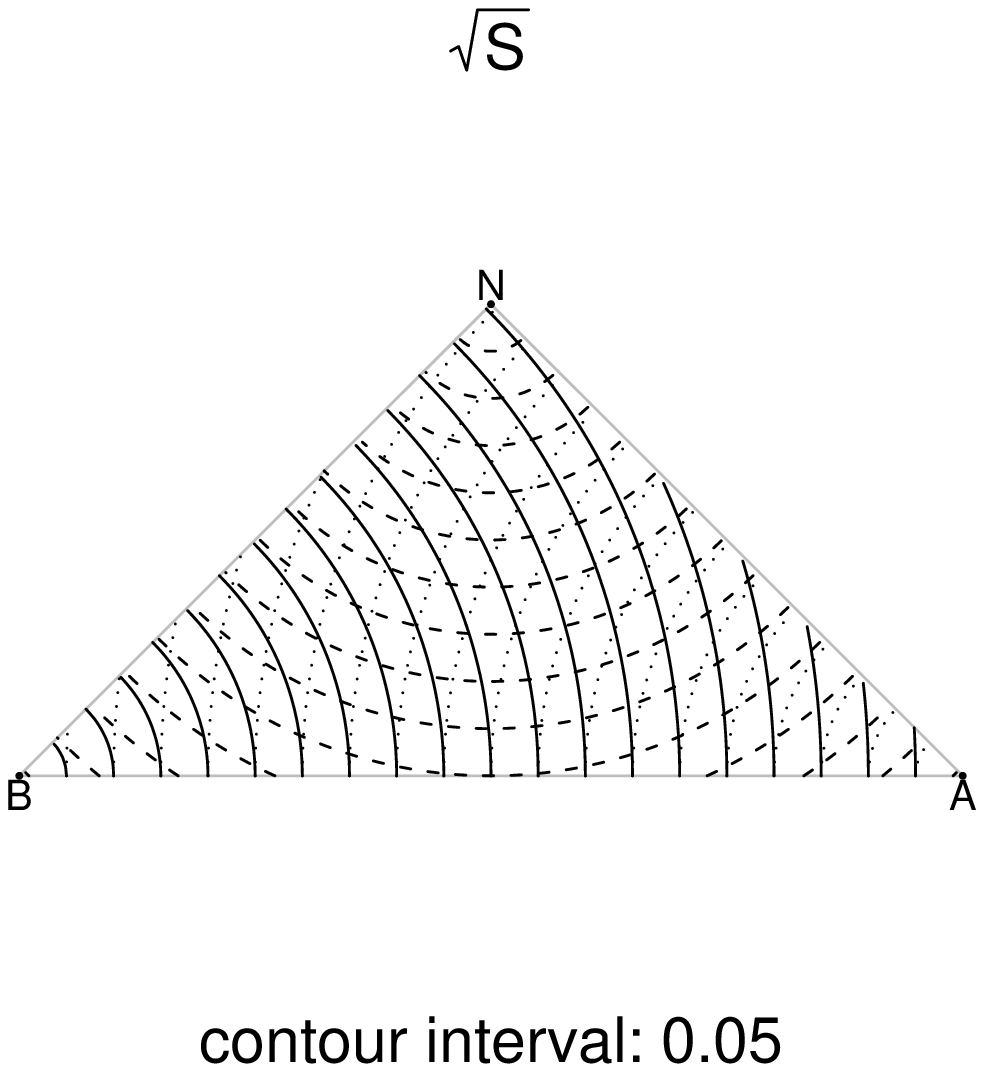}
\caption{Visualisation of ternary forecasts and observations in barycentric coordinates using the categories $B$, $N$ and $A$ defined in Figure \ref{fig:t}. (a) Observations $\mathbf{O}$ can take one of three values and lie at the corners. The angle $\theta$ can be used to compare an arbitrary forecast $\mathbf{P}$ with the climatology $\mathbf{Q}$. (b) A simple way to assign colours to ternary forecasts is to consider the most likely category. (c) For the Brier score, the root--score $\sqrt{S}$ corresponds to distance in an equilateral triangle. Contours of $\sqrt{S}$ are shown for $\mathbf{o}=\mathbf{o}_{B}$ (solid), $\mathbf{o}=\mathbf{o}_{N}$ (dashed), $\mathbf{o}=\mathbf{o}_{A}$ (dotted). (d) as (c) but for the right--angled triangle induced by the Ranked Probability Score.}
\label{fig:Econtours}
\end{center}
\end{figure}

The corners of this triangle are the points $\mathbf{O}_{B} = \hat{M}\mathbf{o}_{B}$, $\mathbf{O}_{N} = \hat{M}\mathbf{o}_{N}$ and $\mathbf{O}_{A} = \hat{M}\mathbf{o}_{A}$ associated with the three possible values of a ternary observation. The point $\mathbf{Q} = \hat{M}\mathbf{q}$ associated with the climatology lies within the triangle. The default choice for the scoring matrix is that associated with the Brier Score, in which case the associated triangle is an equilateral triangle with unit sides. In the case of the default climatology $\mathbf{q}'=(\frac{1}{3},\frac{1}{3},\frac{1}{3})$, the point $\mathbf{Q}$ lies at the centre of this triangle (Figure \ref{fig:Econtours}a). This sort of visualisation receives different names in different disciplines. In mathematics, it is known as a plot in barycentric coordinates while in the applied sciences it is known as a ternary phase diagram.

Barycentric coordinates yield an intuitive geometrical interpretation of ternary forecasts. As well as allowing scores to be visualised as squared distances they allow any scheme for colour assignment to be visualised as a triangular colour palette. For example, Figure \ref{fig:Econtours}b illustrates the algorithm which assigns colours according to which of the three possible outcomes is considered in the forecast to be the most probable. Here and subsequently it is assumed for illustration that the forecast variable $x$ is precipitation. It therefore seems sensible to assign the `dry' colour red to category $B$ and the `wet' colour blue to category $A$. Note that the ternary forecasts $\mathbf{p}'=(1,0,0)$ and $\mathbf{p}'=(0.34,0.33,0.33)$ would both be assigned the colour red in Figure \ref{fig:Econtours} even though they differ greatly. In the former case the forecasting system is \emph{certain} that $x$ will lie in category $B$ while in the latter case the forecast is barely different from the climatology $\mathbf{q}'=(\frac{1}{3},\frac{1}{3},\frac{1}{3})$.

Since scores are squares of distances within the triangle, it is helpful to consider the root--score $\sqrt{S}$. Figure \ref{fig:Econtours}c shows contours of constant root--score in the case of the Brier score, when the observation $\mathbf{o}$ takes each of its three possible values. These contours show the set of forecasts which are equally good (in terms of score) as predictions of the subsequently observed value. Figure \ref{fig:Econtours}d shows contours of constant root--score in the case of the Ranked Probability Score, for which the induced triangle is a right--angled triangle with sides $1/\sqrt{2}$, $1/\sqrt{2}$ and $1$ (Appendix, section \ref{sec:rps}).

\subsection{Current colour schemes for ternary forecasts}\label{sect:oldmethods}

\begin{figure}
\begin{center}
\includegraphics*[angle=0,width=6cm]{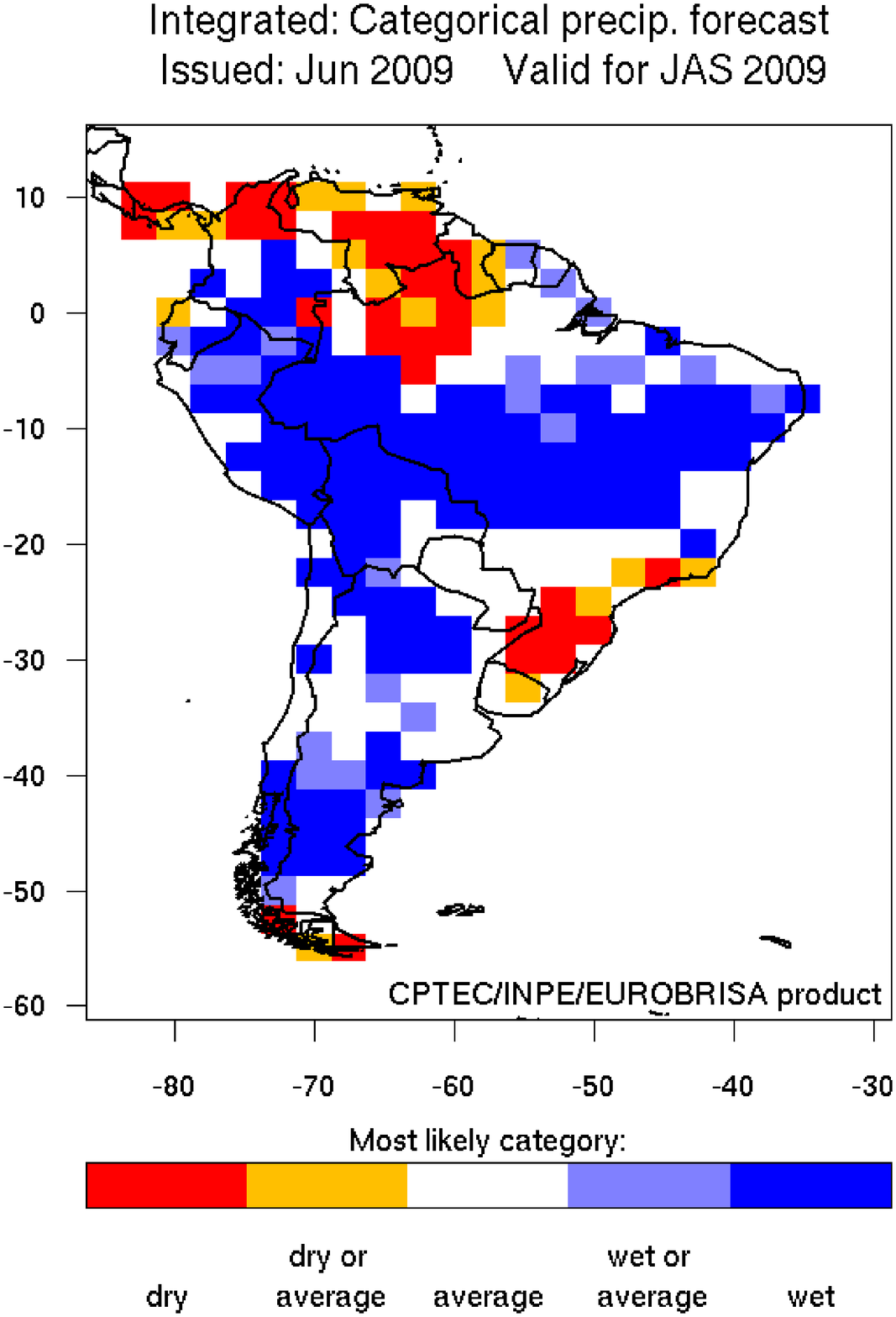}
\includegraphics*[angle=0,width=6cm]{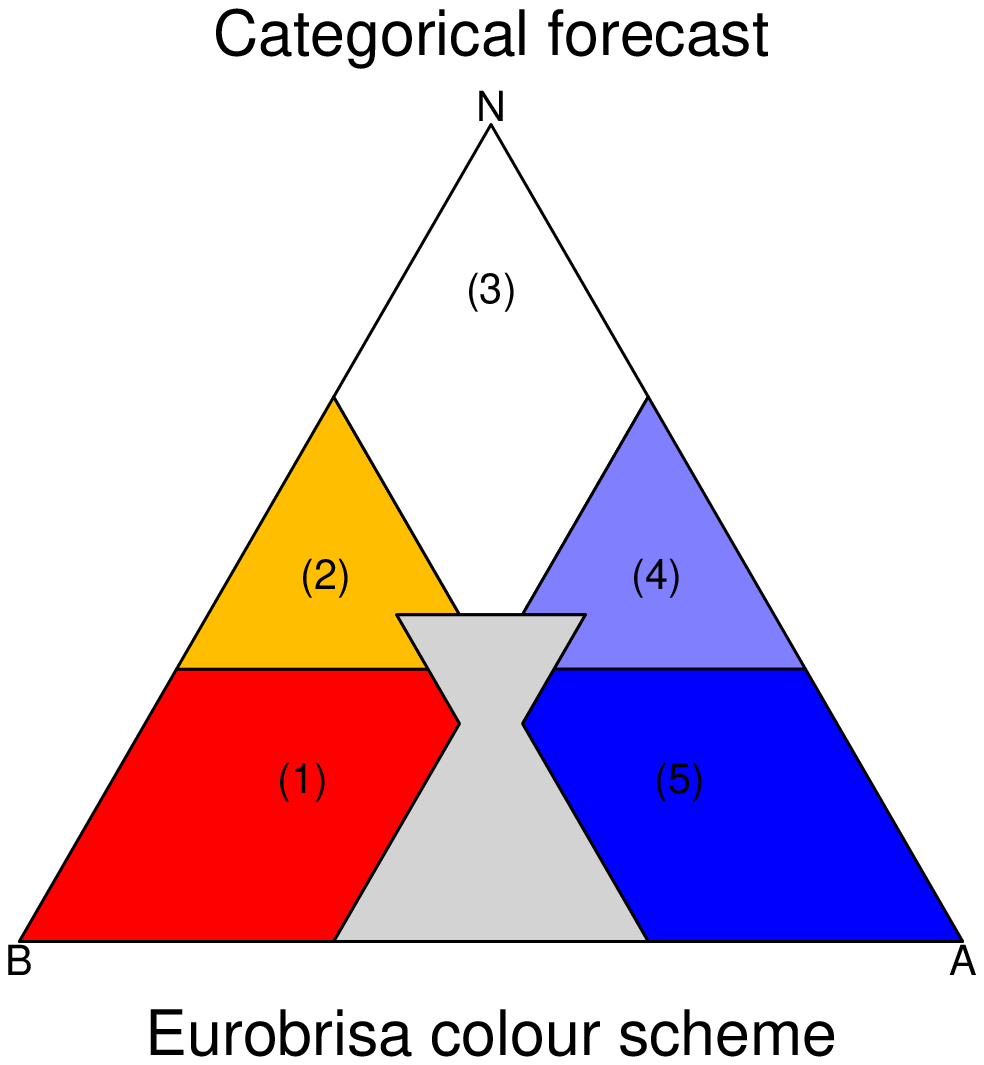}
\caption{(a) Example of a current ternary forecast map (from EUROBRISA). (b) Plotting the colour palette in barycentric coordinates is a useful aid to interpretation.}
\label{fig:euro_cat}
\end{center}
\end{figure}

Current methods for assigning colours to ternary forecasts discretise barycentric coordinates into a finite number of regions, meaning that information present in the forecast is not present in the assigned colour \cite{Slingsby09}. For example, consider the `most likely tercile' scheme of  Figure \ref{fig:Econtours}b used for displaying precipitation forecasts produced in the EURO-BRazilian Initiative for improving South American seasonal forecasts (EUROBRISA) project\footnote{http://eurobrisa.cptec.inpe.br}. Precipitation forecasts produced in EUROBRISA are Gaussian distributions obtained from a calibration and combination procedure known as forecast assimilation \cite{Stephenson05a}. These `integrated' forecasts are a combination of four coupled ocean--atmosphere general circulation models (ECMWF\cite{Anderson07}, UK Met Office \cite{Graham05}, M\'{e}t\'{e}o France \cite{Gueremy05} and CPTEC \cite{Nobre09}) a
nd an empirical model that uses Pacific and Atlantic sea surface temperatures as predictors for South American precipitation \cite{Coelho06}.

A feature of current visualisation methods is that the algorithms by which colours are assigned tend to be described algebraically. In the example considered here, colours are assigned to forecasts based on the following algebraic definitions of 5 regions of forecast space:

1 (Dry):           ($p_B>\frac{2}{5}$ and $p_N<\frac{1}{3}$ and $p_A<\frac{1}{3}$).

2 (Dry or normal): ($p_B>\frac{1}{3}$ and $p_N>\frac{2}{5}$) or ($p_B>\frac{2}{5}$ and $p_N>\frac{1}{3}$).

3 (Normal):        ($p_B<\frac{1}{3}$ and $p_N>\frac{2}{5}$ and $p_A<\frac{1}{3}$).

4 (Wet or normal): ($p_N>\frac{1}{3}$ and $p_A>\frac{2}{5}$) or ($p_N>\frac{2}{5}$ and $p_A>\frac{1}{3}$).

5 (Wet):           ($p_B<\frac{1}{3}$ and $p_N<\frac{1}{3}$ and $p_A>\frac{2}{5}$).

The use of barycentric coordinates allows the meaning of these definitions to be visualised easily (Figure \ref{fig:euro_cat}). It also reveals something that is not immediately obvious from the algebra. There is a region in barycentric coordinates (here coloured grey) that is not included in these definitions. This region of ternary forecast space -- at the base of the triangle -- corresponds to ternary forecasts in which category $N$ is assigned a low probability, but the outlying categories $B$ and $A$ are assigned relatively high probability. Such forecasts can arise when the forecast distribution $F(x)$ has variance much greater than that of the climatology $G(x)$ (\S \ref{sec:normaldist}).

In summary, many current methods for assigning colours to forecasts lose information by discretising ternary forecast space. The same colour is assigned to more than one ternary forecast and the colours do not convey a sense of how much the forecast differs from climatology. A ternary forecast close to climatology provides little gain in information -- the forecast has not told us much that we did not already know. On the other hand, a ternary forecast far from climatology provides a large gain in information. In such a case the forecasting system assigns high probability to the variable lying in one particular category and not the others.

The aim now, therefore, is to assign a continuum of colours to ternary forecasts, viewed as points in barycentric coordinates. This will retain all information present in the ternary forecast and so it will be possible to identify a ternary forecast uniquely from its assigned colour. The choice of colour assignment will, in particular, take account of the information gain in each ternary forecast.

\subsection{Comparing forecasts with the climatology}\label{sect:entropy}

\begin{figure}
\begin{center}
\includegraphics*[angle=0,width=6cm]{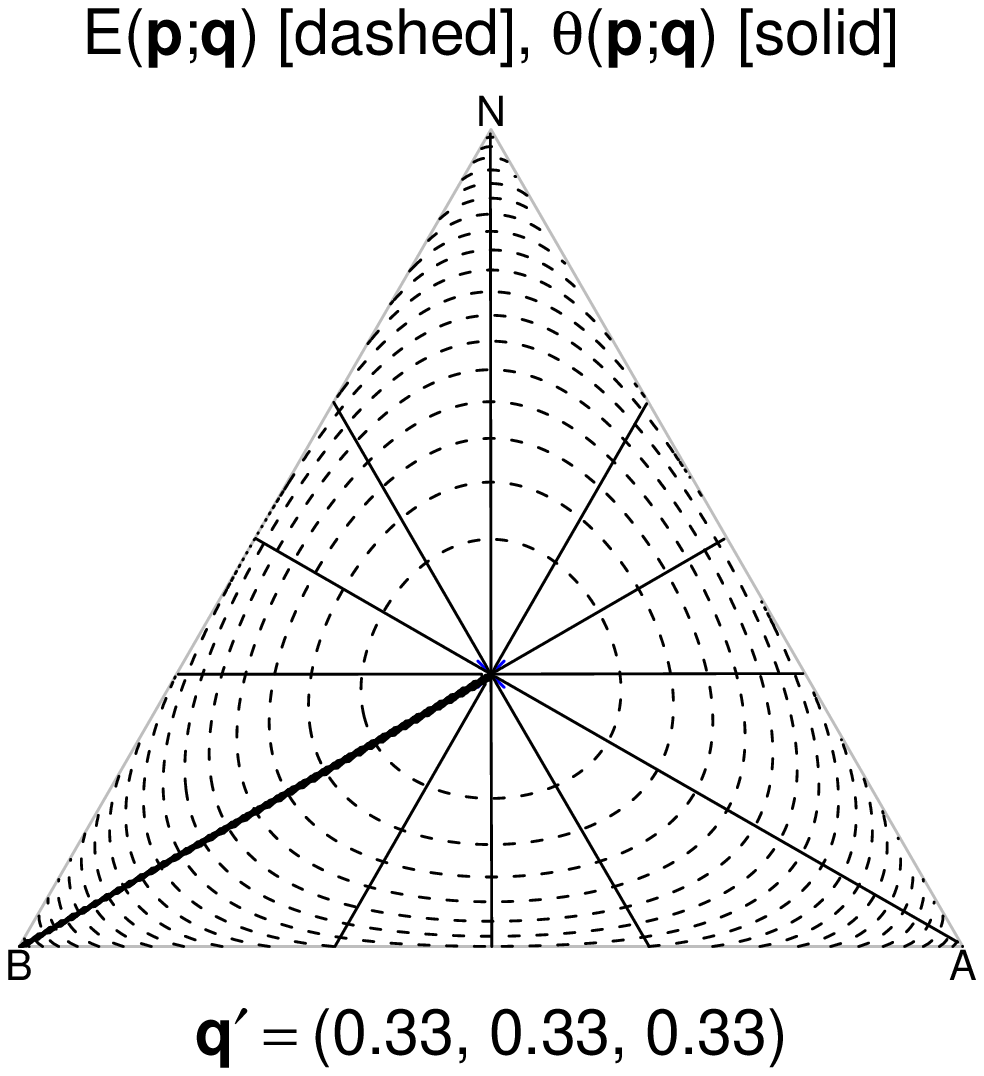}
\includegraphics*[angle=0,width=6cm]{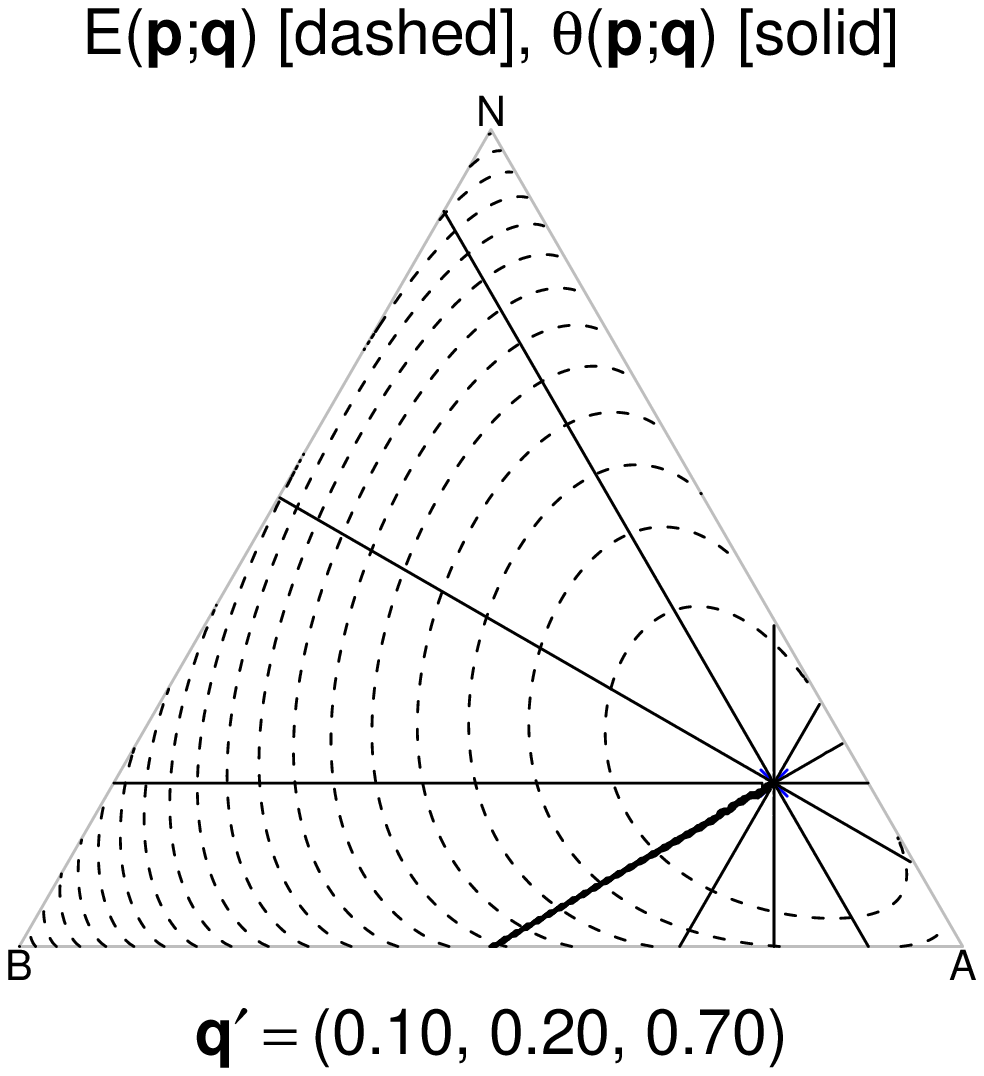}
\caption{(a) Proposed coordinate system in the case when $\mathbf{q}'=(\frac{1}{3},\frac{1}{3},\frac{1}{3})$. Information gain $E(\mathbf{p};\mathbf{q})$ (equation \ref{eq:entropy}) [dashed lines] ranges from $0$ (at the centre) to $1$ (at the corners). Dominant category $\theta(\mathbf{p};\mathbf{q})/2 \pi$ [solid lines] ranges from $0$ to $1$ moving clockwise from thick line. (b) as (a) but with $\mathbf{q}'=(0.1,0.2,0.7)$.}
\label{fig:Scontours}
\end{center}
\end{figure}

Probabilistic forecasts are regarded here as measures of belief and so a measure based on entropy \cite{Jaynes03} is preferred. A sensible measure of information gain $E(\mathbf{p};\mathbf{q})$ is:
\begin{equation}\label{eq:entropy}
E(\mathbf{p};\mathbf{q}) = \left[ \frac{\displaystyle 1}{\displaystyle \log \max_{i} q_{i}^{-1} } \right] \sum_{i \in \{B,N,A \}} p_{i} \log \frac{\displaystyle p_{i}}{\displaystyle q_{i}}
\end{equation}
This can be interpreted as a scaled version of the Kullback--Leibler divergence between $\mathbf{p}$ and $\mathbf{q}$.  Contours of constant $E(\mathbf{p};\mathbf{q})$ are plotted in the unit triangle in Figure \ref{fig:Scontours}. Note that the climatology $\mathbf{q}$ corresponds to an information gain of 0 ($E(\mathbf{q};\mathbf{q})=0$) and that the corner furthest from the climatology corresponds to an information gain of 1 ($E(\mathbf{o};\mathbf{q})=1$ for some $\mathbf{o}$).

Motivated by the idea of assigning colours according to the most likely tercile, the continuous angular measure $\theta(\mathbf{p};\mathbf{q})$ will be referred to as the dominant category. This measures the angle in barycentric coordinates of the forecast $\mathbf{p}$ with respect to an origin at the climatology $\mathbf{q}$. It follows that the information gain $E(\mathbf{p};\mathbf{q})$ and the dominant category $\theta(\mathbf{p};\mathbf{q})$ define an alternative coordinate system for ternary forecast space (Figure \ref{fig:Econtours}).

\subsection{A new colour scheme for ternary forecasts}\label{sect:colours}

In the red--green--blue ($RGB$) representation of colour (used in colour televisions), colours are represented by sets of three numbers in the range $[0,1]$ corresponding to the brightness of the three primary colours. Thus $RGB=(1,0,0)$ is bright red, $RGB=(0,1,0)$ is bright green, and so on. In this paper, colours are assigned using the more intuitive hue--saturation--value ($HSV$) representation. Geometrically, the $RGB$ system defines a unit cube in colour space in which shades of grey occur on a grey--line running from the black corner $RGB=(0,0,0)$ to the white corner $RGB=(1,1,1)$. The $HSV$ system describes colours as points in a cylindrical coordinate system with this grey--line as its axis. In this system, $value \in [0,1]$ is a measure of distance parallel to the grey--line axis, $saturation \in [0,1]$ is a measure of distance perpendicular to the axis and $hue \in [0,1]$ is an angular measure around this axis.
\begin{figure}
\begin{center}
\includegraphics*[width=6cm,angle=0]{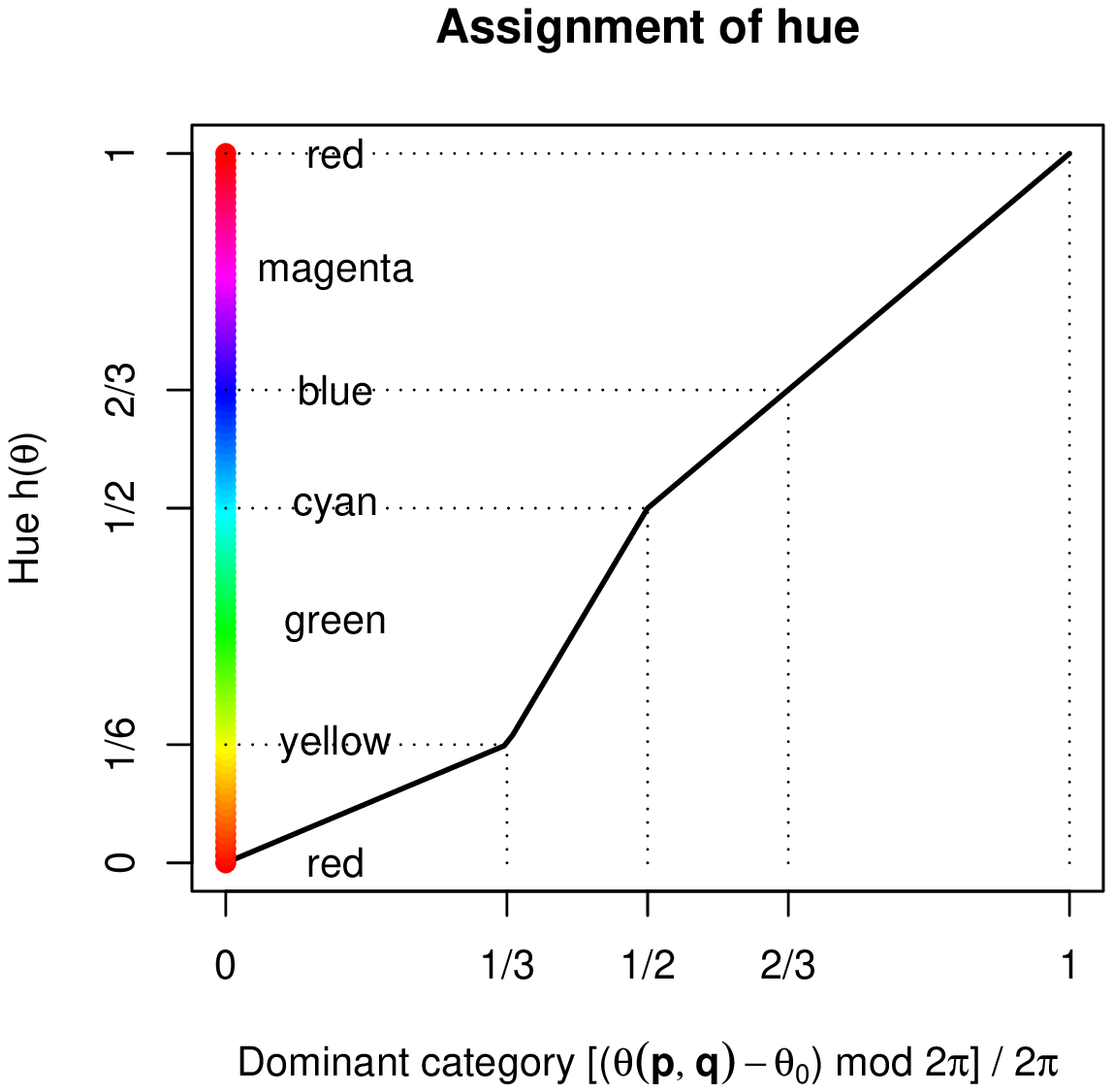}
\includegraphics*[width=6cm,angle=0]{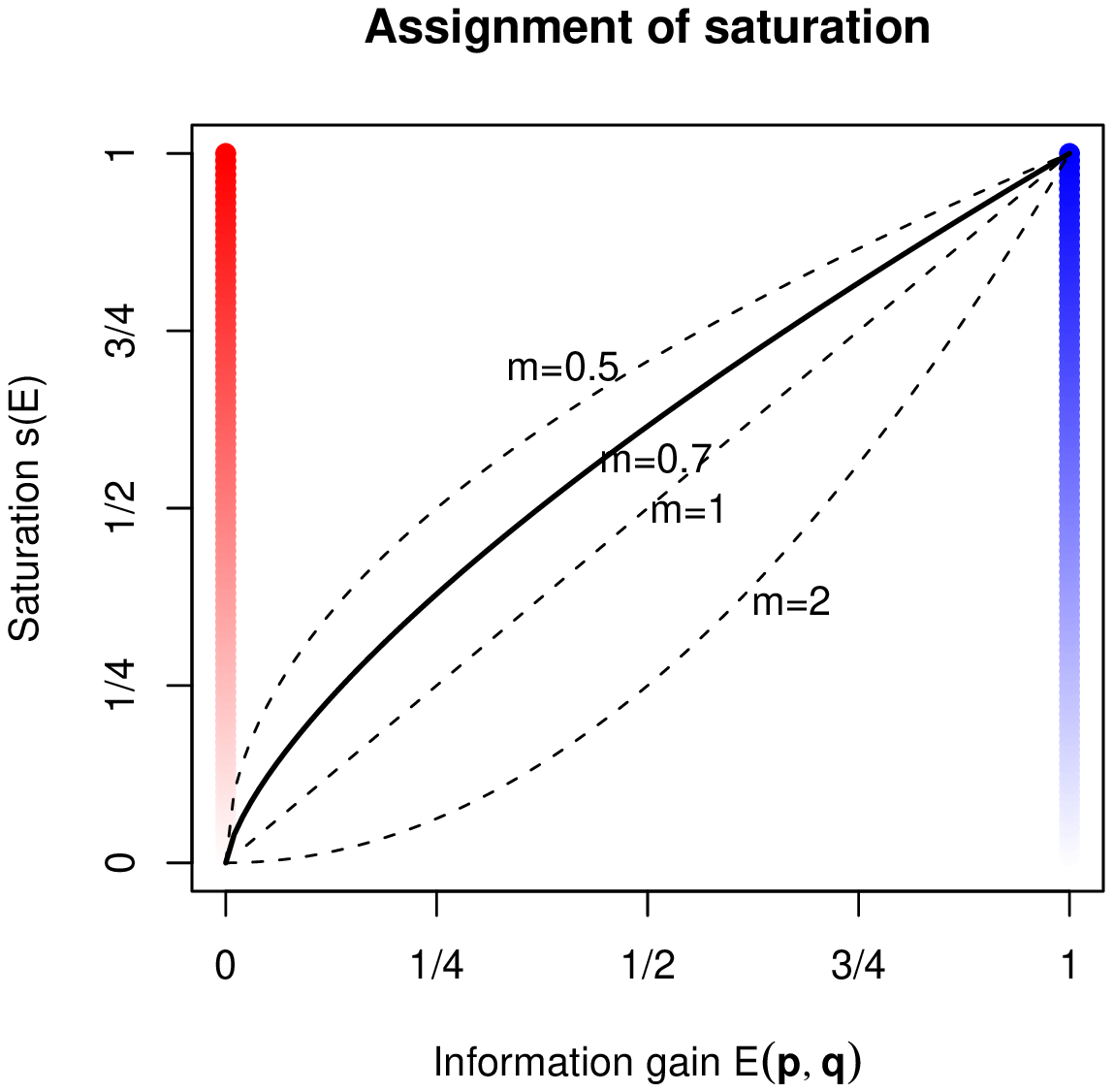}
\caption{The functions of equation \ref{eq:hsv_colours}. (a) Hue as a function of dominant category $\theta(\mathbf{p};\mathbf{q})$. (b) Saturation as a function of information gain $E(\mathbf{p};\mathbf{q})$.  $m=0.7$ and $\theta_{0}=0$ are used throughout this paper.}\label{fig:hue}
\end{center}
\end{figure}

\begin{figure}
\begin{center}
\includegraphics*[angle=0,width=6cm]{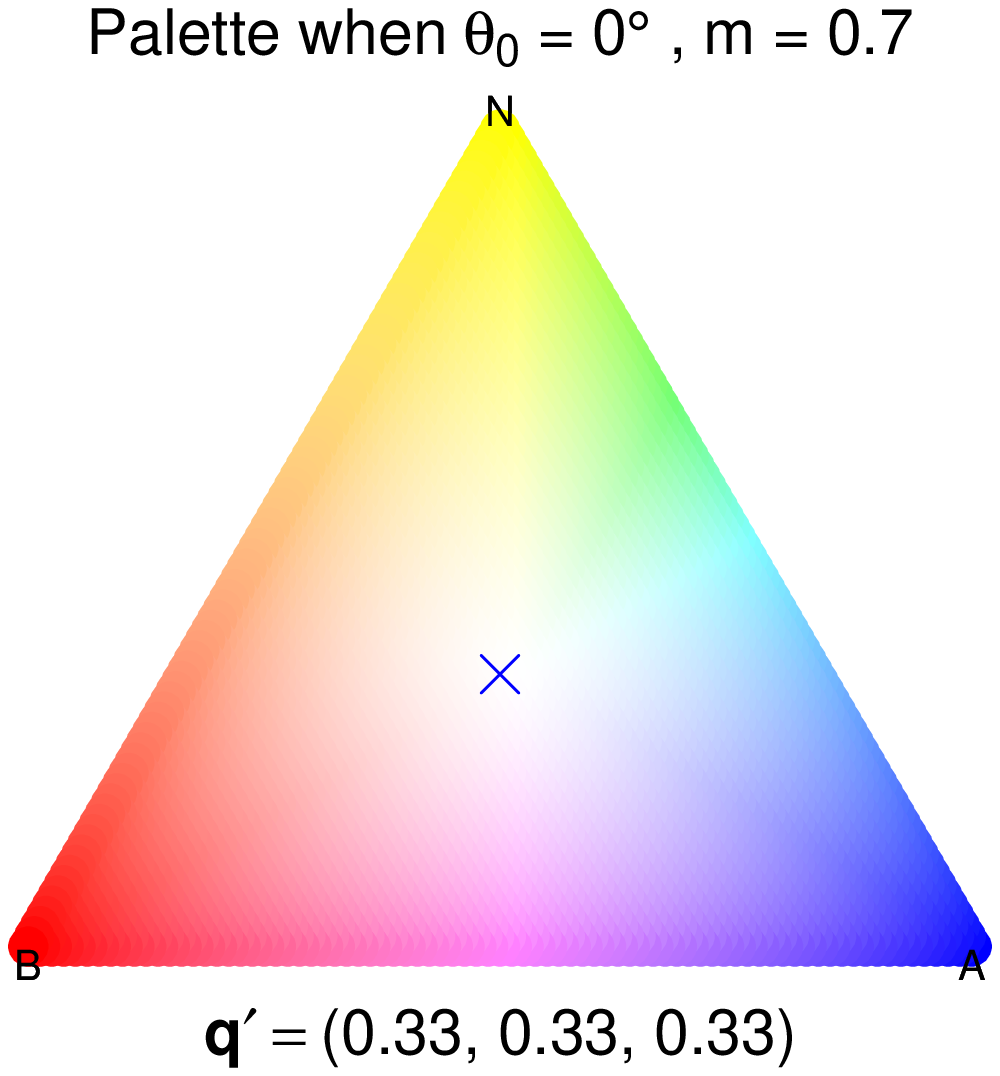}
\includegraphics*[width=6cm,angle=0]{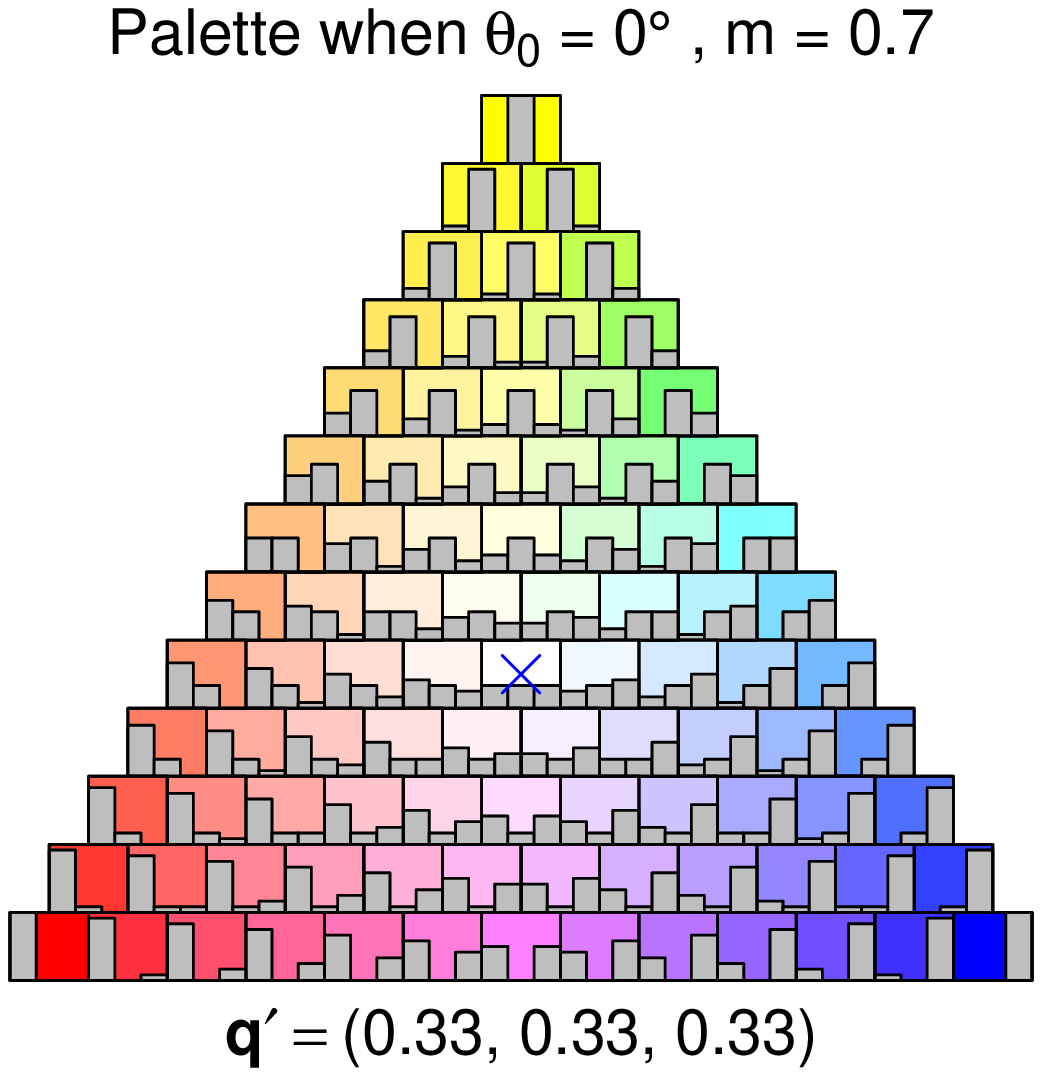}
\caption{(a) Palette of colours assigned to ternary forecasts (equation \ref{eq:hsv_colours}). Climatology $\mathbf{q}$ indicated by blue cross. (b) as (a) but with barplots indicating ternary forecasts $\mathbf{p}$. }\label{fig:bar}
\end{center}
\end{figure}

Here, a colour is assigned to the forecast $\mathbf{p}$ by associating the dominant category $\theta(\mathbf{p};\mathbf{q})$ with the hue and the information gain $E(\mathbf{p};\mathbf{q})$ with the saturation. The proposed algorithm for colour assignment is
\begin{equation}\label{eq:hsv_colours}
\begin{array}{lcl}
\mathrm{hue}         & = & h([ (\theta(\mathbf{p};\mathbf{q}) - \theta_{0}) \mod 2 \pi ] / 2 \pi) \\
\mathrm{saturation}  & = & (E(\mathbf{p};\mathbf{q}))^{m} \\
\mathrm{value}       & = & 1 \\
\end{array}
\end{equation}
where the functional forms $h(\theta)$ and $s(E)=E^m$ are illustrated in Figure \ref{fig:hue}. Suggested default choices for the parameters are $m=0.7$ and $\theta_{0}=0$, which together yield the colour palette shown in Figure \ref{fig:bar}. Note that the climatology is always assigned the colour white by this system. The nonlinear hue function $h(\theta)$ has been chosen to minimise the region of barycentric coordinates that is assigned the colour green and hence to minimise difficulty for readers with green--weak colour blindness \cite{Light04,Stephenson05b}. The parameter $\theta_{0}$ can be chosen in order to rotate the palette in the triangle about the climatology, while the exponent $m$ controls the rate at which colour saturation changes away from the climatology.

\begin{figure}
\begin{center}
\includegraphics*[angle=0,width=12cm]{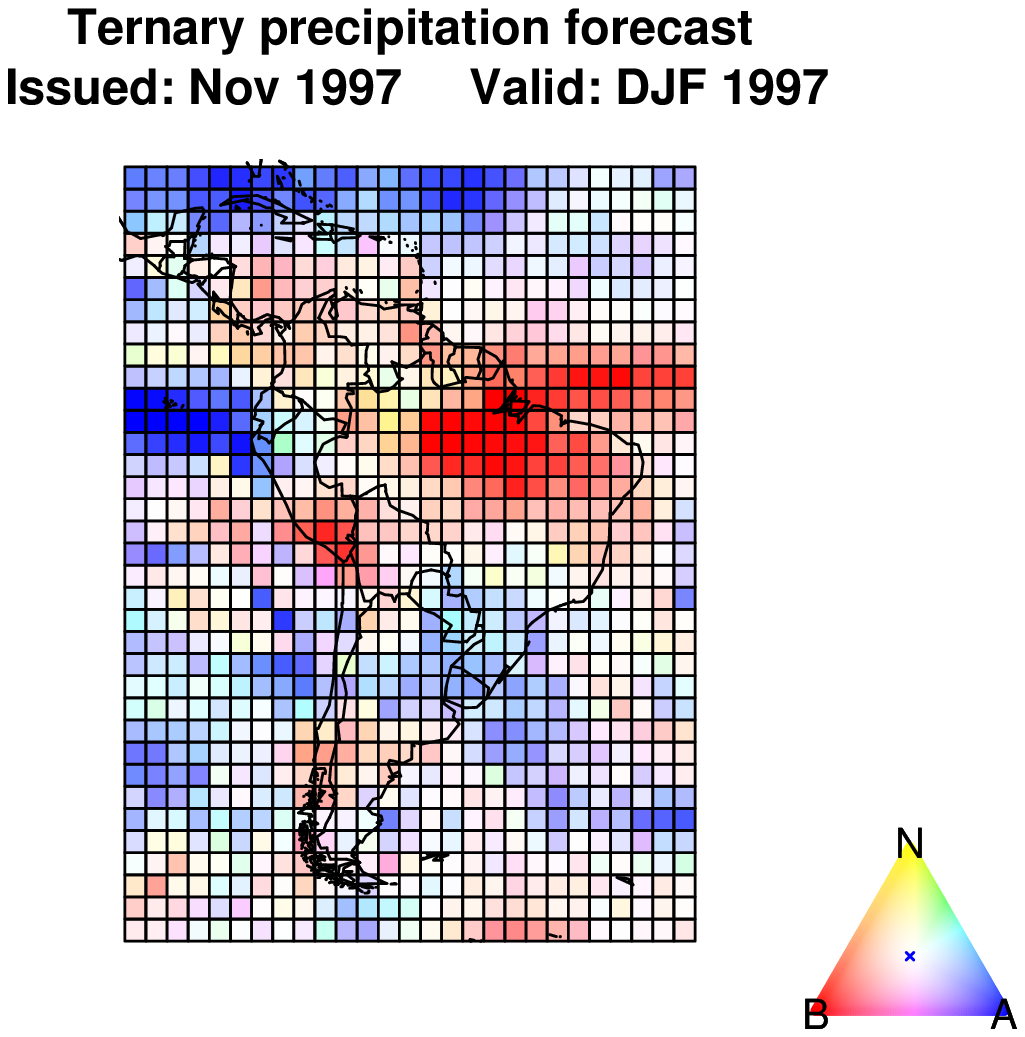}
\caption{A probabilistic forecast map produced using the colour scheme proposed here (equation \ref{eq:hsv_colours}). The continuous colour palette (bottom right) conveys the probabilities assigned to assigned to `Below--', `Near--' and `Above--' normal precipitation categories (Figure \ref{fig:t}). For example, colours close to white indicate a forecast similar to climatology, strong red indicates a high probability of below--normal precipitation, and strong blue a high probability of above--normal precipitation.}
\label{fig:forecast}
\end{center}
\end{figure}

An example of a forecast map using the proposed colour scheme is shown in Figure \ref{fig:forecast}. In this illustration, the data consist of integrated seasonal precipitation forecasts for South America produced by the EUROBRISA project. At each spatial location, the probabilistic forecast is compared with the terciles of the local climatology. In other words, the ternary climatology $\mathbf{q}'=(\frac{1}{3},\frac{1}{3},\frac{1}{3})$ (shown by a cross in the colour palette bottom--right) has been used as the benchmark ternary forecast. The map shows clearly that high probability is assigned to low rainfall in northern Brazil (red) while there are regions in Southeast Brazil and the Southern Ocean for which the forecast barely differs from climatology (white). There are also regions where rainfall in the near--normal category is forecast with high probability (yellow) and low probability (purple).

\subsection{Special case: Gaussian Distributions}\label{sec:normaldist}

To gain insight into the colours assigned by equation \ref{eq:hsv_colours}, it is helpful to consider the special case in which the climatology $G(x)$ and the forecast $F(x)$ are both Gaussian distributions. The space of Gaussian distributions is two--dimensional (each distribution defined by its mean and variance) and so there is a natural one--to--one mapping between a (Gaussian) forecast $F(x)$ and its ternary representation $\mathbf{p}$.

\begin{figure}
\begin{center}
\includegraphics*[angle=0,width=6cm]{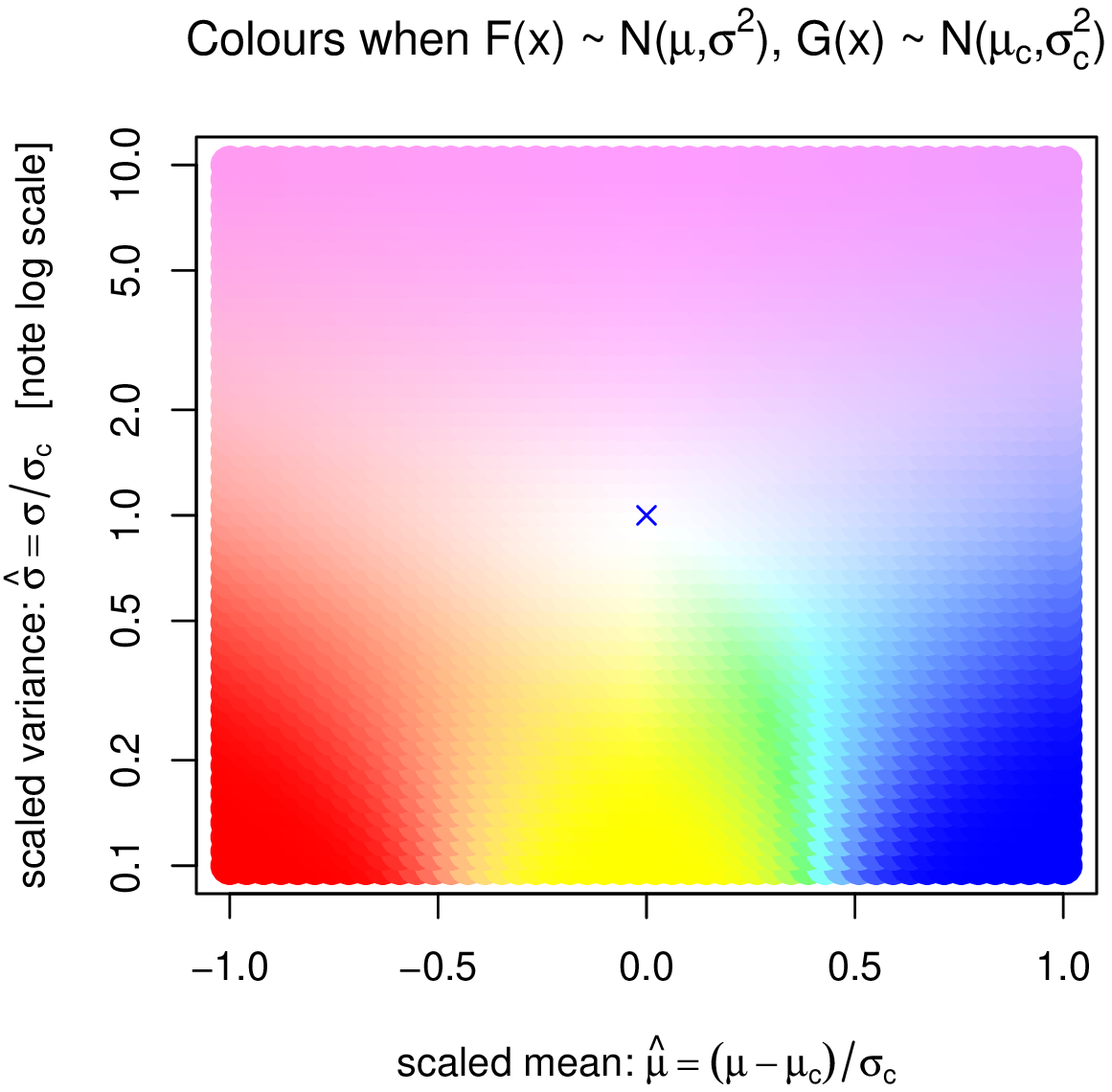}
\includegraphics*[angle=0,width=6cm]{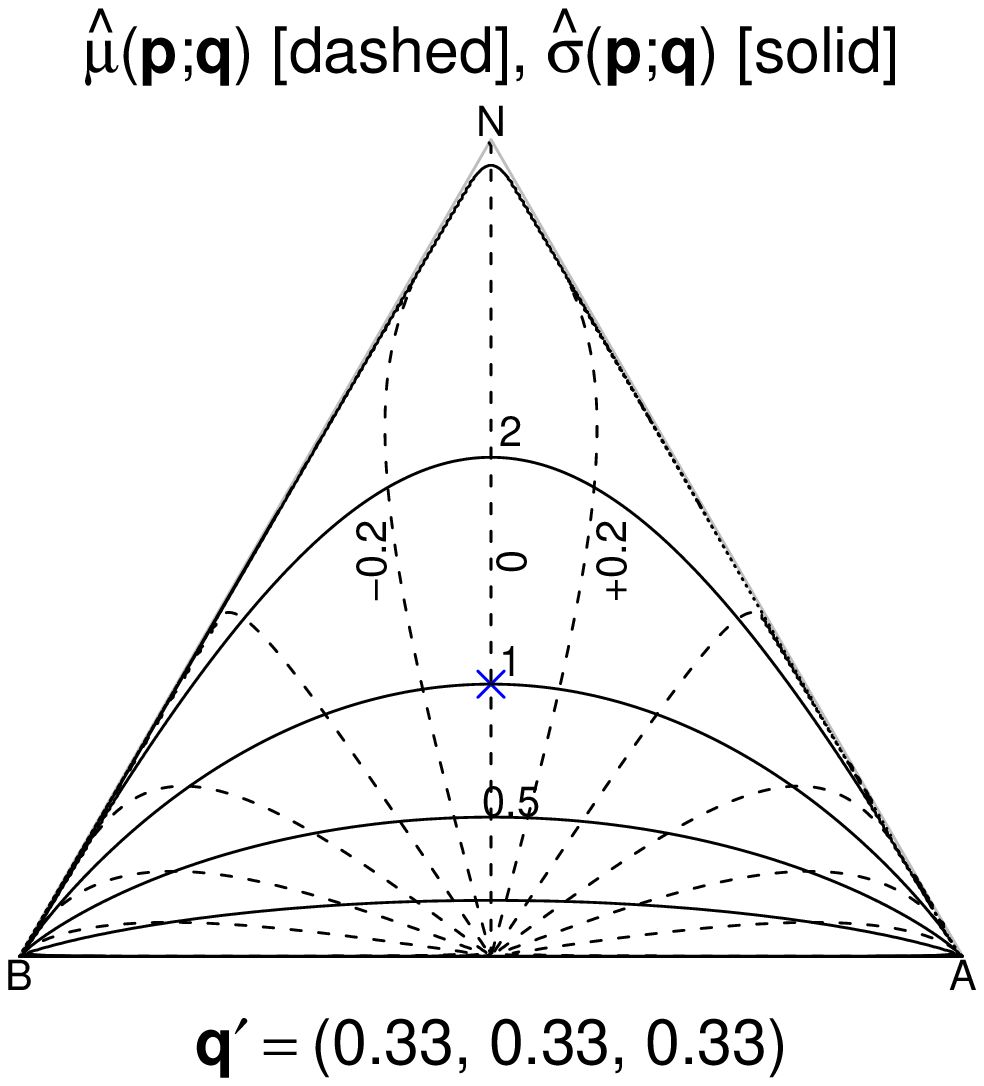}
\caption{(a) Colours assigned when forecast $F(x)$ and climatology $G(x)$ are both Gaussian distributions. The case where the forecast equals the climatology is indicated by a cross. (b) Contours of scaled mean $\hat{\mu} \in \{0, \pm 0.2, \pm 0.5, \pm 1, \pm 2, \pm 5 \}$ and scaled standard deviation $\hat{\sigma}\in \{ 0.2,0.5,1,2,5 \}$ in barycentric coordinates (equation \ref{eq:muhat}).}
\label{fig:musigplot}
\end{center}
\end{figure}

Specifically, suppose that the climatology is $N(\mu_{c},\sigma_{c}^{2})$  and the forecast is $N(\mu,\sigma^{2})$. It follows that
\begin{equation}
G(x) = \Phi \left(\frac{x-\mu_{c}}{\sigma_{c}} \right) ; \quad F(x) = \Phi \left(\frac{x-\mu}{\sigma} \right)
\end{equation}
where $\Phi(z)$ is the CDF of the standard normal distribution $N(0,1)$. It is helpful to normalise the forecast mean and standard deviation with respect to those of the climatology
\begin{equation}\label{eq:muhat}
\hat{\mu} = \frac{\mu-\mu_{c}}{\sigma_{c}} ; \quad \hat{\sigma} = \frac{\sigma}{\sigma_{c}}
\end{equation}
It follows that the forecast is more sharply peaked than the climatology when $\hat{\sigma} < 1$ and less sharply peaked than the climatology when $\hat{\sigma} > 1$.
The ternary forecast $\mathbf{p}$ can be calculated from equation \ref{eq:q} by noting that
\begin{equation}\label{eq:gauss}
F(x_{B}) = \Phi \left( \frac{\Phi^{-1}(q_{B})-\hat{\mu}}{\hat{\sigma}} \right) ; \quad
F(x_{A}) = \Phi \left( \frac{\Phi^{-1}(q_{B}+q_{N})-\hat{\mu}}{\hat{\sigma}} \right)
\end{equation}

Figure \ref{fig:musigplot}a illustrates the colours that are assigned to a forecast and climatology when both are Gaussian. When the forecast is much more sharply peaked than the climatology ($\hat{\sigma} \ll 1$) the ternary forecast assigns nearly all of its probability mass to one of the three categories. When the forecast mean is much less than the climatological mean ($\hat{\mu} \ll 0$) this probability is assigned to category $B$ and the assigned colour is a strong red. Similarly a strong yellow is assigned when the means are comparable ($\hat{\mu} \approx 0$) and a strong blue is assigned when the forecast mean significantly exceeds the climatological mean ($\hat{\mu} \gg 0$)

The situation is rather different when the forecast has much higher variance than the climatology ($\hat{\sigma} \gg 1$). In this case the ternary forecast assigns little probability mass to the central category $N$ but rather splits it approximately evenly between the two extreme categories. In this case the assigned colour is purple. This sort of situation might arise in an ensemble climate forecast in which some members of the ensemble predict that the climate variable increases while other members predict that it decreases. The purple colour might also arise if the forecasting system predicted an increase in the frequency of extreme events in both tails of the distribution.

Figure \ref{fig:musigplot}b shows $\hat{\mu}$ and $\hat{\sigma}$ in barycentric coordinates, and hence the Gaussian distributions of climatology and forecast that would yield given a ternary forecast $\mathbf{p}$. Consider the case in which the forecast and the climatology have similar variances, as appears often to be the case. It follows that $\hat{\sigma} \approx 1$  and so the ternary forecast $\mathbf{p}$ lies close to the contour $\hat{\sigma} \approx 1$ in Figure \ref{fig:musigplot}b. The forecast mean $\mu$ then controls the colour assigned to the forecast, with red assigned when $\hat{\mu} \ll 1$, white assigned when $\hat{\mu} \approx 1$ and blue assigned when $\hat{\mu} \gg 1$.

\section{Verification of ternary probabilistic forecasts}\label{sec:verify}

The previous sections have considered the visualisation of a set of probabilistic forecasts. The aim in this section is to visualise the difference between a set of forecasts and the corresponding set of observations. This is known as forecast verification \cite{Jolliffe03}. In the next section the standard score decomposition of \cite{Murphy73} is reviewed and re--interpreted geometrically.

\subsection{Decomposition of the score}

Suppose that the scoring matrix $L$ has been specified so that the score of an individual forecast--observation pair is given by a squared distance in the appropriate triangle (equation \ref{eq:scorein R2}). As in previous sections, the Brier score is the default choice but the results hold for any quadratic scoring function.

The triangle representing ternary forecast space can be discretised into a finite number of bins with centres $\mathbf{P}_{k}$. For simplicity, each forecast $\mathbf{P}$ that lies in the bin with centre $\mathbf{P}_{k}$ will be reassigned the central value $\mathbf{P}_{k}$ \cite{Doblas08}.  Following \cite{Murphy73}, consider the identity
\begin{equation}
\mathbf{P}_{k}-\mathbf{O} = (\mathbf{P}_{k}-\mathbf{\overline{O}}|\mathbf{P}_{k}) - (\mathbf{Q}-\mathbf{\overline{O}}|\mathbf{P}_{k}) + (\mathbf{Q}-\mathbf{O})
\end{equation}
where $\mathbf{\overline{O}}|\mathbf{P}_{k}$ is the mean observation associated with forecasts in bin $\mathbf{P}_{k}$. If the climatology is defined to be the mean of all the observations $\mathbf{Q} = \mathbf{\overline{O}}$, it follows that a decomposition of the mean score with binned forecasts is:
\begin{equation}\label{eq:decomp}
\begin{array}{ccccccc}
\overline{ \| \mathbf{P}_{k} - \mathbf{O} \|^{2} } &
= &
\overline{ \| \mathbf{Q} - \mathbf{O} \|^{2} } &
- &
\overline{ \| \mathbf{Q} - \mathbf{\overline{O}|P}_{k} \|^{2} } &
+ &
\overline{ \| \mathbf{P}_{k} - \mathbf{\overline{O}|P}_{k} \|^{2} }
 \\
\mathrm{score} &
= &
\mathrm{uncertainty} &
- &
\mathrm{resolution} &
+ &
\mathrm{reliability} \\
S &
= &
U &
- &
Z &
+ &
R
\end{array}
\end{equation}
Equation \ref{eq:decomp} constitutes the standard decomposition of the score $S$ into uncertainty $U$, reliability $R$ and resolution $Z$ \cite{Jolliffe03}. It is important to stress that this decomposition applies only when all forecasts within a bin are assigned the central value \cite{Stephenson08}.

\begin{figure}
\begin{center}
\includegraphics*[angle=0,width=6cm]{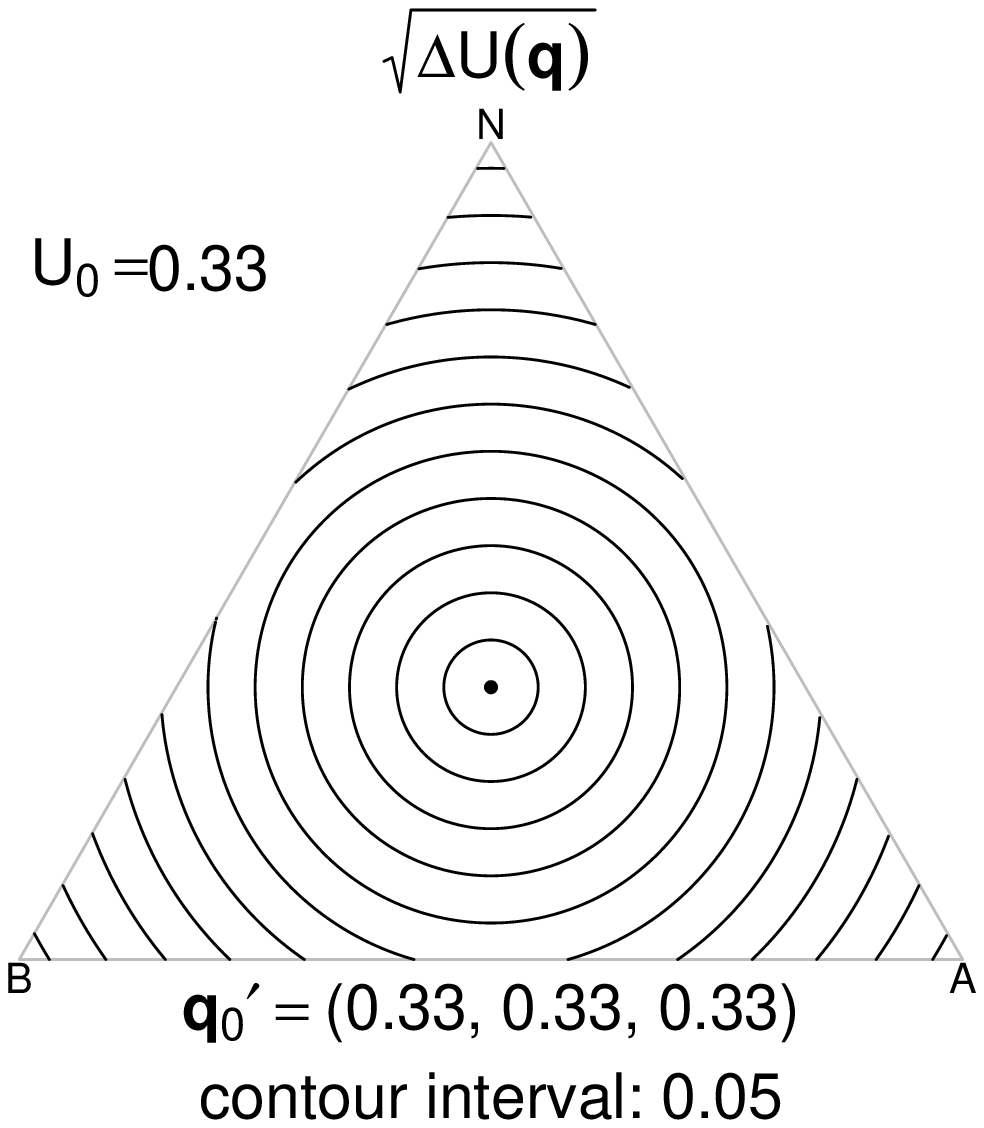}
\includegraphics*[angle=0,width=6cm]{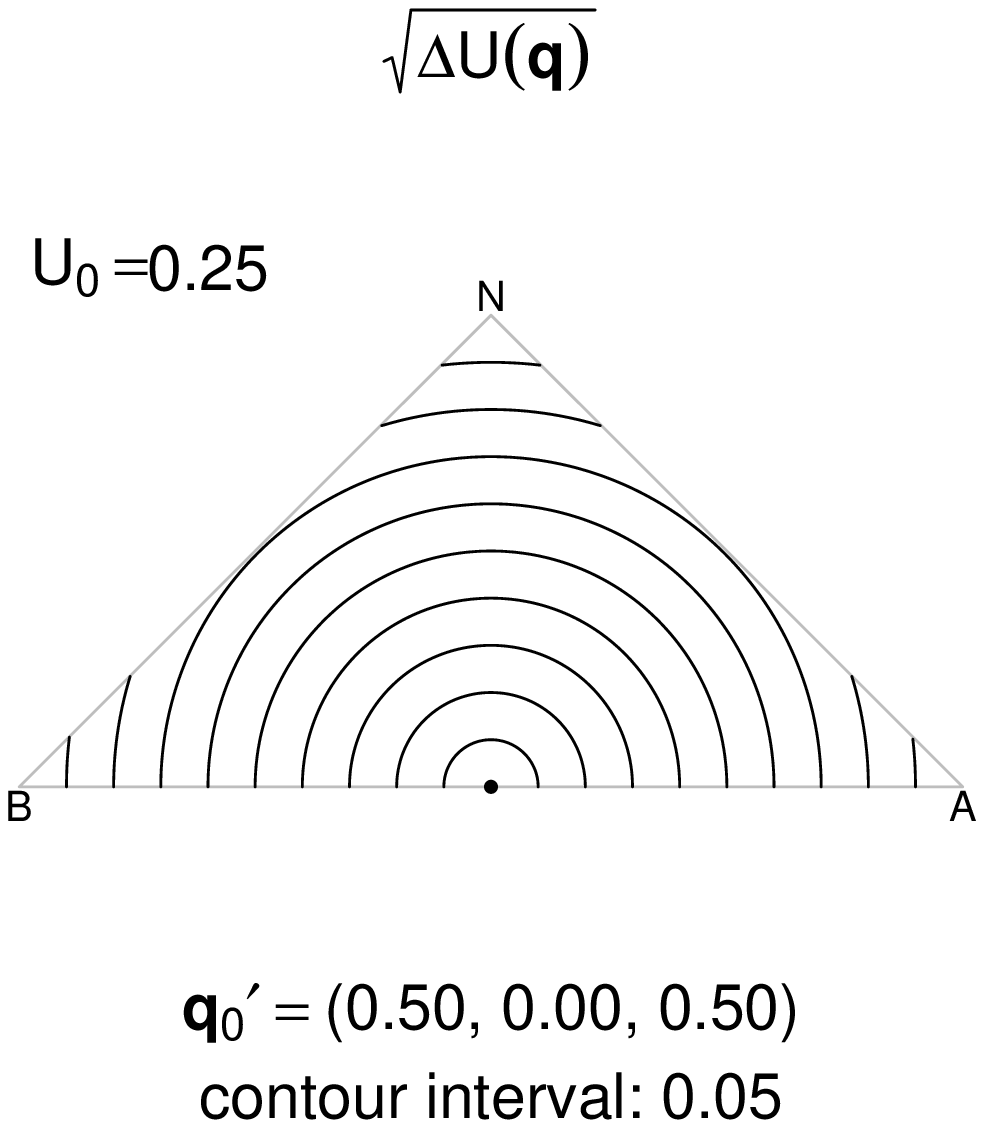}
\caption{The root-uncertainty--reduction $\sqrt{\Delta U}$ (equation \ref{eq:U}) for the Brier score (left) and Ranked Probability Score (right) interpreted as distances within the appropriate triangles.}
\label{fig:Ucontours}
\end{center}
\end{figure}

It has already been remarked that the score is the mean square distance in the triangle between forecasts and observations, and it is clear that the uncertainty is the mean square distance between observations and climatology. In order to see how uncertainty varies with climatology $\mathbf{q}$ (Figure \ref{fig:Ucontours}), it is helpful (Appendix, section \ref{sec:Uappendix}) to consider the particular climatology $\mathbf{q}_{0}$ (shown by a dot in Figure \ref{fig:Ucontours}) for which the uncertainty gains its maximum value $U_{0} = U(\mathbf{q}_{0})$. The uncertainty $U(\mathbf{q})$ is then
\begin{equation}\label{eq:U}
U = U_{0} - \Delta U
\end{equation}
where $\Delta U (\mathbf{q})$ represents the reduction in uncertainty between $\mathbf{q}_{0}$ and $\mathbf{q}$. Figure \ref{fig:Ucontours} illustrates that the root--uncertainty--reduction $\sqrt{\Delta U}$ can be visualised as the distance in the triangle between the climatology of interest $\mathbf{q}$ and the climatology $\mathbf{q}_{0}$ of maximum uncertainty.

Similarly, the resolution in equation \ref{eq:decomp} is the mean square distance between climatology $\mathbf{Q}$ and the mean observations conditional on the forecasts $\mathbf{\overline{O}|P}_{k}$. The reliability is the mean square distance between forecasts $\mathbf{P}_{k}$ and the mean observations conditional on the forecasts $\mathbf{\overline{O}|P}_{k}$.

\begin{figure}
\begin{center}
\includegraphics*[angle=0,width=10cm]{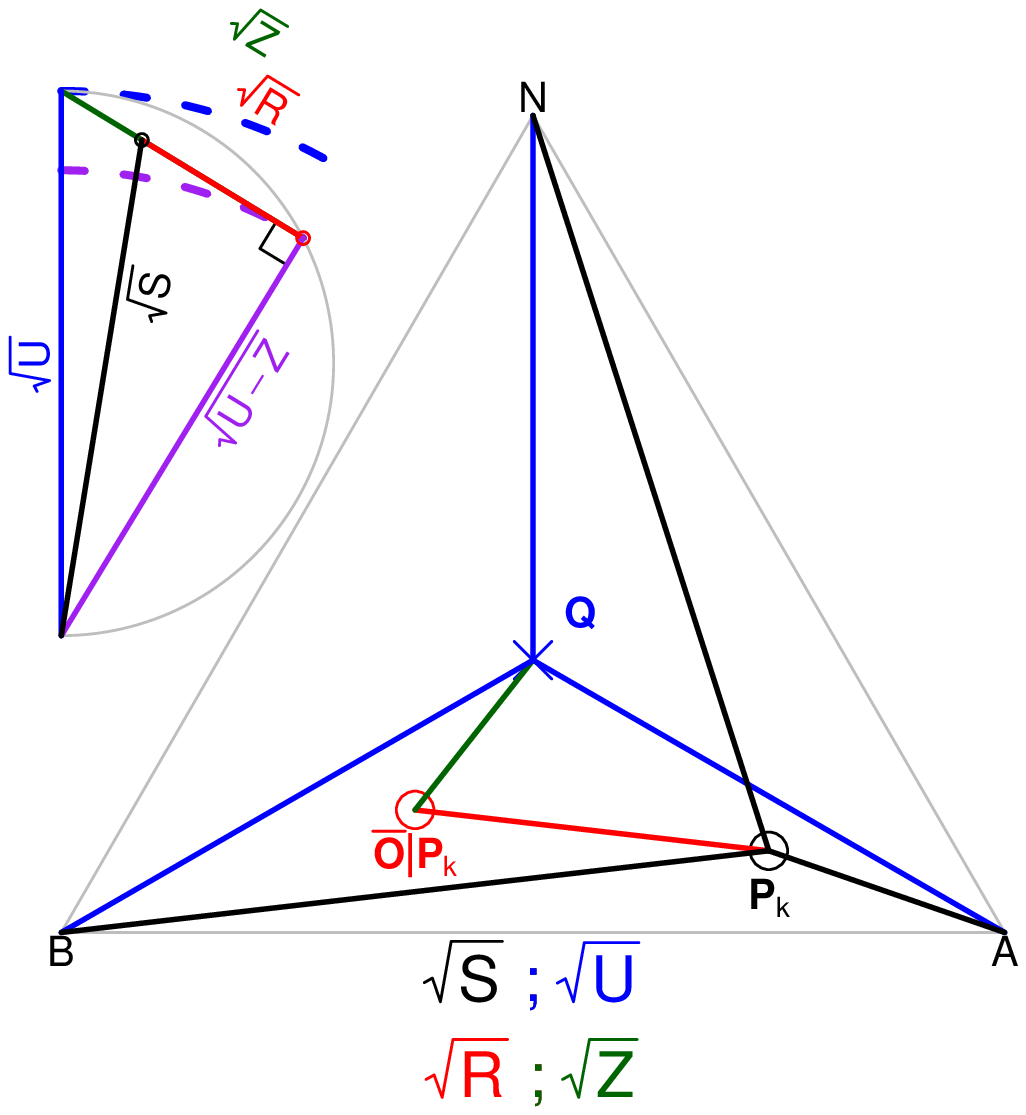}
\caption{Geometrical interpretation of the decomposition of a quadratic score $S$ into uncertainty $U$, resolution $Z$ and reliability $R$ (equation \ref{eq:decomp}), whose square--roots represent distances in the triangle. A proposed 'decomposition diagram' is shown at top--left.}
\label{fig:decomp}
\end{center}
\end{figure}

It follows that the root--score $\sqrt{S}$ is the root--mean--square (rms) distance between forecasts and observations, the root--uncertainty $\sqrt{U}$ is the rms distance between climatology and observations, the root--resolution $\sqrt{Z}$ is the rms distance between climatology and mean conditional observations and the  root--reliability $\sqrt{R}$ is the rms distance between forecasts and mean conditional observations. These geometrical interpretations as rms distances are illustrated in the triangle in Figure \ref{fig:decomp}.

The aim of verification is to quantify the root--score $\sqrt{S}$ (Figure \ref{fig:decomp}, black lines) and the aim of recalibration is to minimise the root--score. Since the forecaster cannot change the observations (Figure \ref{fig:decomp}, red circles), recalibration proceeds by changing the forecasts $\mathbf{P}_{k}$ (Figure \ref{fig:decomp}, black circles) in order to minimise the root--reliability $\sqrt{R}$ (Figure \ref{fig:decomp}, red lines).

The diagram at the top left of Figure \ref{fig:decomp} contains a proposed graphical interpretation of the score decomposition. The root--uncertainty $\sqrt{U}$ is independent of the forecasts and defines a semi--circle (grey) of diameter $\sqrt{U}$ (blue). The root--resolution $\sqrt{Z}$ is also independent of the value of the forecasts (provided the `binning' of the forecasts remains unchanged) and defines the larger of the two right--angled triangles. This triangle has sides of length $\sqrt{U}$ (blue), $\sqrt{Z}$ (green) and $\sqrt{U-Z}$ (purple). It follows from Pythagoras' theorem that all three corners of this triangle lie on the semi--circle, with two on the diameter. The smaller of the two right--angled triangles contains components of the score decomposition that can be altered by recalibration of the forecasts. The smaller triangle has sides of length $\sqrt{S}$ (black), $\sqrt{R}$ (red) and $\sqrt{U-Z}$ (purple) and equation \ref{eq:decomp} constitutes Pythagoras' theorem. The aim of recalibration is to minimise the root--score $\sqrt{S}$ by minimising the root--reliability $\sqrt{R}$. It follows that the dashed lines in the decomposition diagram represent limiting values of the root--score $\sqrt{S}$. A forecasting system that always issued the climatology as its forecast would have resolution $Z=0$ and hence have a root--score $\sqrt{S}$ indicated by the blue dashed line. This represents the worst performance that might be expected from a skilful forecasting system. On the other hand, the best possible performance is indicated by the purple dashed line. This would occur for a forecasting system with perfect root--reliability $\sqrt{R}=0$.

\subsection{Ternary reliability diagrams}

\begin{figure}
\begin{center}
\includegraphics*[angle=0,width=6cm]{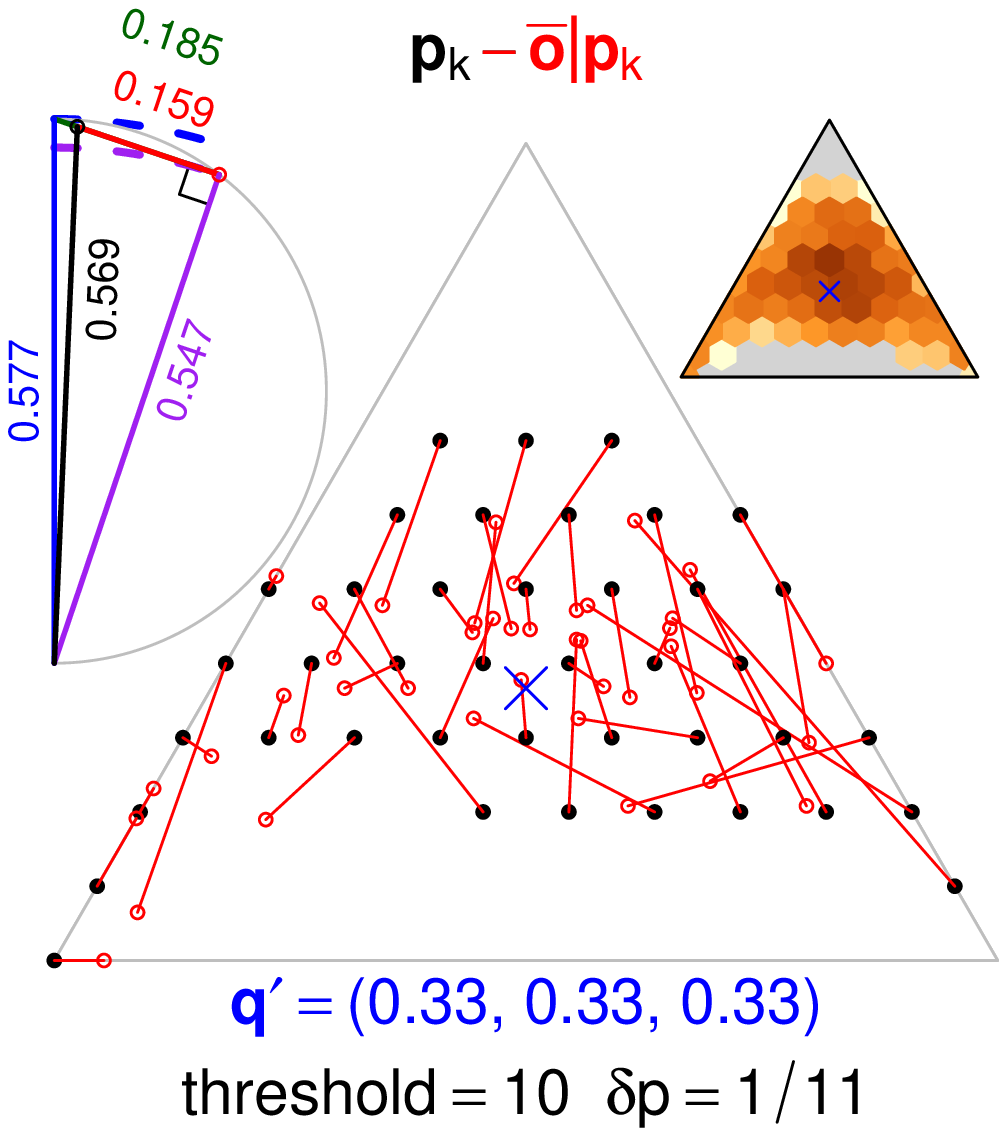}
\includegraphics*[angle=0,width=6cm]{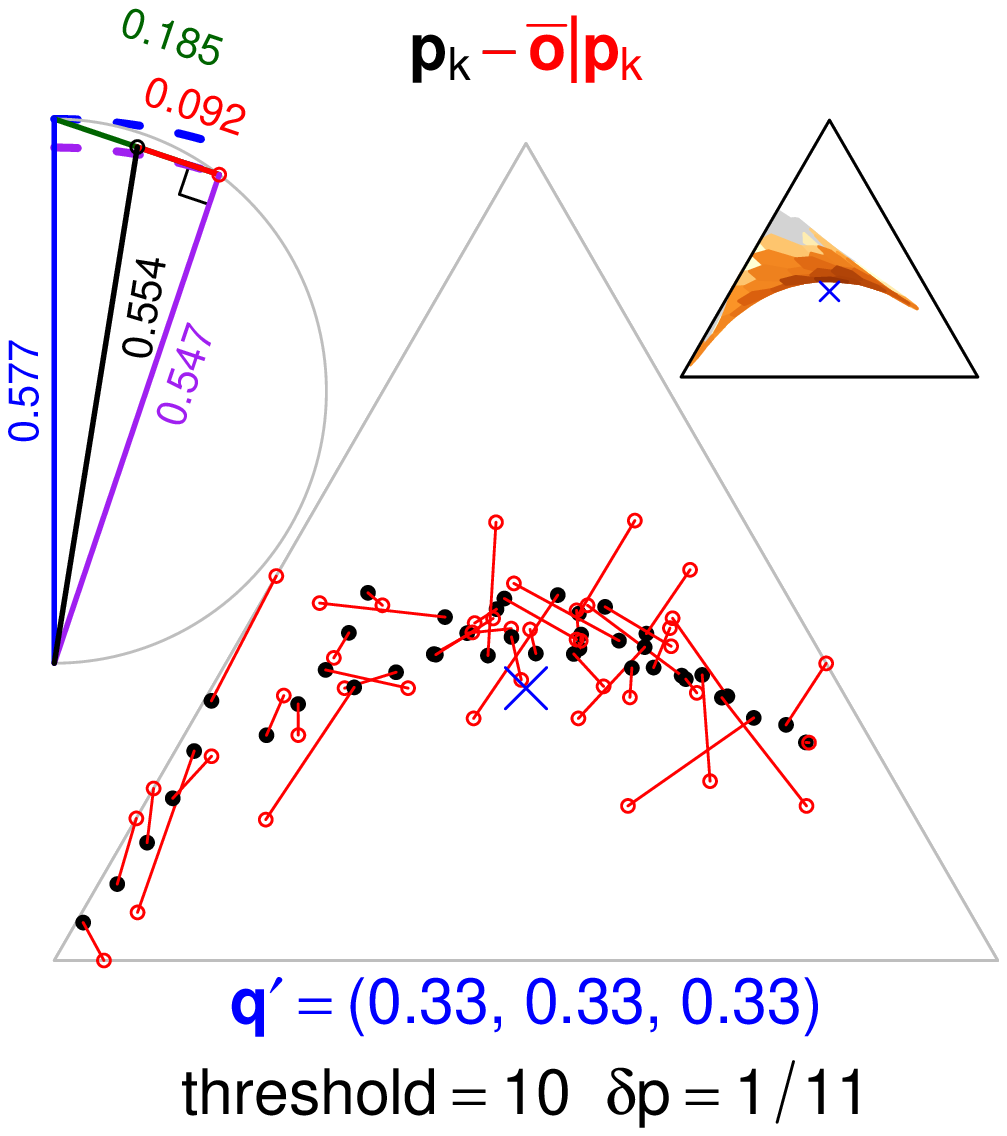}
\caption{Examples of proposed `ternary reliability diagrams' (a) A set of poorly calibrated `original' forecasts has poor reliability (long red dipoles). (b) Recalibration of these forecasts by `moving the black dots' improves reliability (shorter red dipoles). Decomposition diagram (top left) shows that recalibrated score is close to the best achievable by recalibration. Sharpness diagram (top right) illustrates the recalibration geometrically.}
\label{fig:cal}
\end{center}
\end{figure}

The ideas of the preceding section can now be illustrated with an example. The data in this example are seasonal precipitation hindcasts at one month lead--time from the EUROBRISA integrated forecast model. The data relate to the years 1981 -- 2005, the season January--February--March (JFM) and a spatial region $72.5^{\circ}W-42.5^{\circ}W$, $12.5^{\circ}S-2.5^{\circ}N$ for which seasonal forecasts show reasonable skill.

Figure \ref{fig:cal}a contains one of our proposed ternary reliability diagrams. The large triangle contains information on the reliability of the forecasts. Bins of size $\delta \mathbf{p} = \frac{1}{11}$ were chosen and the central forecast in each bin $\mathbf{P}_{k}$ -- to which all forecasts in that bin are set equal -- is plotted in the triangle as a black circle. The mean observation conditional on a forecast $\mathbf{\overline{O}|P}_{k}$ is plotted as a red circle and the two are joined by a red line line in order to form a `dipole'. The length of each dipole represents the root--reliability of the forecasts in each bin. The number of forecasts in each bin is shown graphically by the ternary sharpness diagram at the top--right. (Dark colours -- high density, light colours -- low density, grey -- no data). The climatology is shown by a blue cross. Dipoles for bins containing fewer than 10 forecasts (indicated in the diagram by the text `threshold =10') are omitted for clarity. An advantage of this sort of visualisation of the data is that coherent patterns in the reliability dipoles are immediately obvious. Such patterns would indicate regions of ternary forecast space in which the forecasting system consistently assigned the wrong probability to one of the three categories.

The decomposition diagram (top--left) illustrates geometrically the decomposition of the root--score in this dataset. Thus, the root--score $\sqrt{S}=0.569$ is seen to be rather poor compared to the best--possible root--score $\sqrt{U-Z}=0.547$ that could be attained through recalibration.

\subsection{Forecast maps including verification data}

\begin{figure}
\begin{center}
\includegraphics*[angle=0,width=12cm]{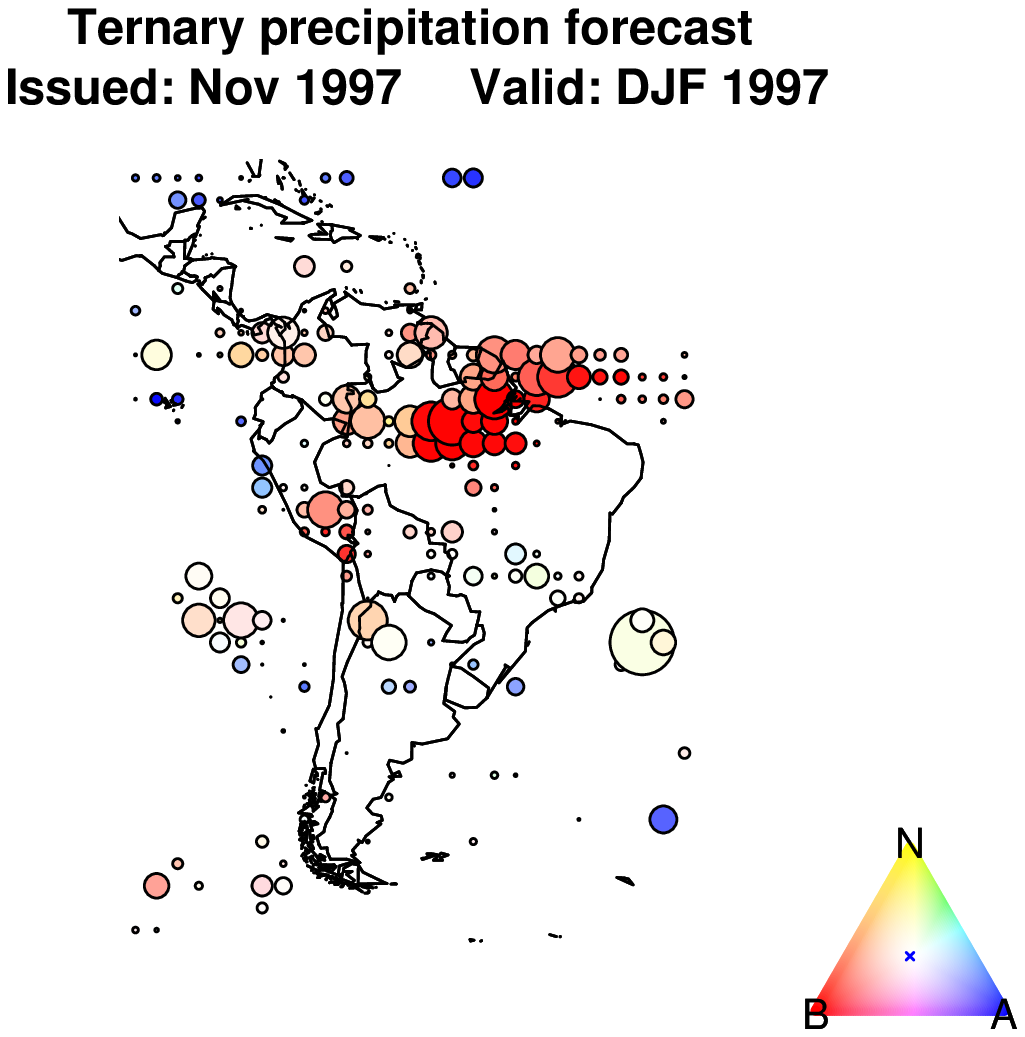}
\caption{Probabilistic forecast map including verification data. Larger circles correspond to forecasting system performing well historically. No circle is plotted where forecasting system has performed worse than a climatological forecast. Colours assigned as in figure \ref{fig:forecast}, circle radius via equation \ref{eq:radius}.  }
\label{fig:forecast-circles}
\end{center}
\end{figure}

The map of Figure \ref{fig:forecast} contains information about forecasts but not about the skill of the forecasting system. In order to incorporate verification data into a forecast map, the array of coloured squares can be replaced by an array of coloured circles whose radii are a measure of forecast skill in a verification data set. One possibility would be to set radius proportional to $(Z-R)/Z$ which is a `standard' definition of skill. In this paper, however, the convention has been to consider quantities like $\sqrt{Z}$ and $\sqrt{R}$ which can be interpreted as distances in a triangle rather than squared distances (Figure \ref{fig:decomp}) and so $(\sqrt{Z}-\sqrt{R})/\sqrt{Z}$ is preferred here. (It is straightforward to relate these two possibilities via the identity $(Z-R)/Z=1-(1-(\sqrt{Z}-\sqrt{R})/\sqrt{Z})^{2}$.) Making the choice:
\begin{equation}\label{eq:radius}
\mathrm{circle \ radius} \propto \frac{\sqrt{Z}-\sqrt{R}}{\sqrt{Z}}
\end{equation}
it follows that larger circles will be plotted in regions of high skill, and no circle will be plotted if the forecasting system has less skill than a climatological forecast (i.e. when the assigned radius is negative). A map produced in this way is shown in Figure \ref{fig:forecast-circles}. This map shows clearly that, for the season shown, the skill of this forecasting system is greatest near the northern coast of South America.  Maps of this type should prove useful in communication of operational forecasts because the reader's eye is draw to areas where the forecasting system has been shown to perform well.

\section{Recalibration of ternary probabilistic forecasts}\label{sec:calib}

The decomposition diagram in Figure \ref{fig:cal}a illustrates that significant improvement can sometimes be made to the mean--score by recalibrating the forecasts. The score of the original forecasts is $\overline{ \|\mathbf{P}-\mathbf{O} \|^{2} }$ and so the aim is to produce recalibrated forecasts $\tilde{\mathbf{p}}$, calculated from the original forecasts via some specified functional form, whose score $\overline{ \|\tilde{\mathbf{P}}-\mathbf{O} \|^{2} } $ is minimised. As an example, consider recalibrated ternary forecasts which are a quadratic function of the original ternary forecasts (Appendix, section \ref{sec:nlf}). Given a dataset of forecasts and observations, standard numerical optimisation techniques can be used to find coefficients in the quadratic function which minimise the score of the recalibrated forecasts. The recalibration function can be interpreted by comparing Figure \ref{fig:cal}a and Figure \ref{fig:cal}b. Recalibration minimises the score by changing the reliability (that is, the root--mean--square length of the red dipoles). It is important to note that recalibration changes the forecasts (black dots) but not the observations (red circles) and so the rms length of the red--dipoles has been reduced from $\sqrt{R}=0.159$ to $\sqrt{R}= 0.092$ solely by moving the position of black dots. The effect of this reduction can be seen in the decomposition diagrams (top--left). Recalibration changes neither the root--uncertainty $\sqrt{U} = 0.577$ nor the root--resolution $\sqrt{Z} = 0.185$, but reduces the root--score from $\sqrt{S} = 0.569$ to $\sqrt{S}=0.554$.

In this case it is immediately clear that the recalibrated forecasts (Figure \ref{fig:cal}b, black dots) are clustered around an arc passing close to the climatology (blue cross). It is interesting that this arc is similar to the contour $\hat{\sigma}=1$ in Figure \ref{fig:musigplot}b which applies when the climatology and the forecast are both Gaussian with equal variance.

Finally, the sharpness diagram (Figure \ref{fig:cal}b, top--right) shows the distribution of the recalibrated forecasts. It is produced by passing the sharpness diagram of the original forecasts (Figure \ref{fig:cal}a, top--right) through the quadratic recalibration function (Appendix section \ref{sec:nlf}). Regions of the sharpness diagram coloured grey indicate possible forecasts that were not originally. Regions coloured white indicate forecasts that can never be issued under this recalibration.

\section{Conclusions}\label{sec:conclusions}

This article has outlined a novel procedure for visualising ternary probabilistic forecasts. The proposed maps convey through colour all of the information present in a ternary forecast. A colour palette printed next to the map can be used to deduce the ternary forecasts directly from the colour used.

The proposed colour scheme always assigns the colour white to a forecast close to climatology, and strong colours to forecasts that differ greatly from climatology. The examples considered in this paper have concerned precipitation and so the `wet' colour blue has been assigned to the above normal category $A$ and the `dry' colour red to the below normal category $B$. It is straightforward to reverse these conventions if desired for other forecast variables such (e.g. temperature). The default palette (with $\theta_{0}=0$) assigns strong red, yellow and blue respectively to forecasts that assign high probability to one of the three categories. Yellow has been chosen in preference to green in order to assist colour blind readers. A variety of palettes can be created by varying the parameters $m$ and $\theta_0$ in equation \ref{eq:hsv_colours}.

A novel visual interpretation of ternary forecasts has also been suggested for verification and recalibration under quadratic scoring functions. It has been shown that Brier scores correspond to squared distances in an equilateral triangle and Ranked Probability Scores correspond to squared distances in a right--angled triangle. Thus, the root--score, root--uncertainty, root--resolution and root--reliability can all be interpreted as root--mean square distances. Verification data can be visualised using the proposed ternary reliability diagrams (e.g. Figure \ref{fig:cal}), which incorporate a decomposition diagram (top--left) and a sharpness diagram (top--right) to aid interpretation. The geometrical interpretation can also be applied to nonlinear recalibration of probabilistic forecasts.

It is hoped that the procedures outlined here will prove useful in the interpretation and communication of operational weather and climate forecasts. When no information on skill is available a map like that of Figure \ref{fig:forecast} can be produced. If skill information is available, it can be  incorporated as in Figure \ref{fig:forecast-circles}.

\section*{Acknowledgements}

This work was supported by the EUROBRISA network project (F/00 144/AT) kindly funded by the Leverhulme Trust. The dynamical ensemble forecast data were kindly provided by ECMWF as part of the EUROSIP project. Three forecasting centres are the partners in EUROSIP, these are ECMWF, the UK Met Office and M\'{e}t\'{e}o-France. CASC was supported by Funda\c{c}\~{a}o de Amparo \`{a} Pesquisa do Estado de S\~{a}o Paulo (FAPESP), processes 2005/05210-7 and 2006/02497-6. RL and DBS thank Aidan Slingsby of City University for useful discussions.

\appendix{}\label{seq:app}

\begin{figure}
\begin{center}
\includegraphics*[width=5cm,angle=0]{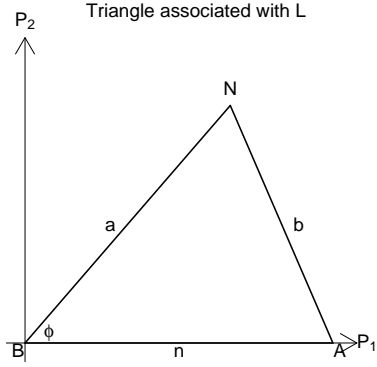}
\caption{The triangle in $\mathbb{R}^{2}$ induced by the score matrix $L$. }\label{fig:tri}
\end{center}
\end{figure}

Let the triangle in $\mathbb{R}^{2}$ induced by the matrix $L$ have vertices $BNA$ (Figure \ref{fig:tri}) and sides of length $b$, $n$ and $a$:
\begin{equation}
b = \sqrt{S(\mathbf{o}_{N};\mathbf{o}_{A})}, \
n = \sqrt{S(\mathbf{o}_{A};\mathbf{o}_{B})}, \
a = \sqrt{S(\mathbf{o}_{B};\mathbf{o}_{N})}
\end{equation}
From the cosine rule, the angle $\phi$ is given by
\begin{equation}
\cos \phi = \frac{n^2+a^2-b^2}{2an}
\end{equation}
It follows that a ternary forecast $\mathbf{p} \in \mathbb{R}^{3}$ and the associated point $\mathbf{P} \in \mathbb{R}^{2}$ within the induced triangle are related by:
\begin{equation}\label{eq:transforms}
\mathbf{P} = \hat{M} \mathbf{p}; \quad \mathbf{p} = M \mathbf{P} + \mathbf{o}_{B}
\end{equation}
where the transformation matrices are defined by
\begin{equation}\label{eq:M}
\hat{M}  = \left(
       \begin{array}{ccc}
         0 & a \cos \phi & n \\
         0 & a \sin \phi & 0
       \end{array}
     \right) ; \quad
M  = \frac{1}{an \sin \phi}\left(
       \begin{array}{cc}
         -a \sin \phi & a \cos \phi -n \\
         0 & n \\
         a \sin \phi & -a \cos \phi
       \end{array}
     \right)
\end{equation}

\subsection{Brier Score}\label{sec:brier}

The Brier Score is defined by
\begin{equation}
\begin{array}{rcl}
S(\mathbf{p};\mathbf{o}) & = &  \frac{1}{2}\left[ (p_B - o_B)^{2} + (p_N - o_N)^{2} + (p_A - o_A)^{2} \right]
\end{array}
\end{equation}
It follows that, for the Brier Score,
\begin{equation}\label{eq:usualr3r2}
L = \frac{1}{\sqrt{2}}\left(\begin{array}{ccc}
                              1 & 0 & 0 \\
                              0 & 1 & 0 \\
                              0 & 0 & 1
                            \end{array}
\right); \quad
\hat{M}  = \left(
       \begin{array}{ccc}
         0 & \frac{1}{2} & 1 \\
         0 & \frac{\sqrt{3}}{2} & 0 \\
       \end{array}
     \right) ; \quad
M  = \left(
       \begin{array}{ccc}
         -1 & \frac{-1}{\sqrt{3}} \\
         0 & \frac{2}{\sqrt{3}} \\
         1 & \frac{-1}{\sqrt{3}} \\
       \end{array}
     \right)
\end{equation}
The triangle induced by the Brier Score is an equilateral triangle with unit sides $b=n=a=1$.

\subsection{Ranked Probability Score}\label{sec:rps}

The Ranked Probability Score \cite{Epstein69, Murphy69, Murphy71} is defined by:
\begin{equation}\label{eq:R}
S(\mathbf{p};\mathbf{o}) =  \frac{1}{2}\left[ (p_B - o_B)^{2} + (p_B+p_N - o_B-o_N)^{2} \right]
\end{equation}
It follows that, for the Ranked Probability Score,
\begin{equation}
L = \frac{1}{\sqrt{2}}\left(\begin{array}{ccc}
                              1 & 0 & 0 \\
                              1 & 1 & 0 \\
                              1 & 1 & 1
                            \end{array}
\right); \quad
\hat{M}  = \left(
       \begin{array}{ccc}
         0 & \frac{1}{2} & 1 \\
         0 & \frac{1}{2} & 0 \\
       \end{array}
     \right) ; \quad
M  = \left(
       \begin{array}{ccc}
         -1 & -1 \\
         0  & 2 \\
         1  & -1 \\
       \end{array}
     \right)
\end{equation}
The triangle induced by the Ranked Probability Score is a right--angled triangle with sides $b=a=1/\sqrt{2}$, $n=1$.

\subsection{The uncertainty of a quadratic score}\label{sec:Uappendix}
For a quadratic scoring rule  (equation \ref{eq:score}) the uncertainty $U(\mathbf{q})$ is equal to the expected score when the climatology $\mathbf{q}$ is issued as the forecast. It follows that
\begin{equation}
U(\mathbf{q}) = \mathbf{v}'\mathbf{q} - \mathbf{q}'L'L\mathbf{q};\  \mathrm{where} \ \mathbf{v} = \mathrm{diag}(L'L)
\end{equation}
It follows that
\begin{equation}
U(\mathbf{q}) = \frac{1}{2}\mathbf{q}_{0}'\mathbf{v} - (\mathbf{q}-\mathbf{q}_{0})'L'L(\mathbf{q}-\mathbf{q}_{0}) ; \ \mathrm{where} \
\mathbf{q}_{0}=\frac{1}{2}(L'L)^{-1}\mathbf{v}
\end{equation}
In the particular cases of the Brier and Ranked Probability Scores:
\begin{equation}\label{eq:R_of_q}
\overline{U} = \left\{
\begin{array}{ll}\frac{1}{2}(1 - \mathbf{q}'\mathbf{q}) & \mathrm{(Brier \ score)} \\ \frac{1}{2}(q_{B}(1-q_{B})+q_{A}(1-q_{A})) & \mathrm{(Ranked \ probability \ score)}
                                                 \end{array}\right.
\end{equation}

\subsection{Quadratic recalibration}\label{sec:nlf}
An original forecast $\mathbf{p}'=(p_{B},p_{N},p_{A})$ is defined by $p_{B}$ and $p_{A}$ since $p_{N}=1-p_{B}-p_{A}$. Define a recalibrated forecast $\tilde{\mathbf{p}}'=(\tilde{p}_{B},\tilde{p}_{N},\tilde{p}_{A})$
by
\begin{equation}
\begin{array}{rcl}
  \tilde{p}_{B} & = & C_{1} + C_{2} p_{B} + C_{3} p_{A} + C_{4} p_{B}^{2} + C_{5} p_{B}p_{A} + C_{6} p_{A}^{2} \\
  \tilde{p}_{A} & = & C_{7} + C_{8} p_{B} + C_{9} p_{A} + C_{10} p_{B}^{2} + C_{11} p_{B}p_{A} + C_{12} p_{A}^{2} \\
  \tilde{p}_{N} & = & 1 - \tilde{p}_{B} - \tilde{p}_{A}
\end{array}
\end{equation}
Numerical values for the parameters $C_{1}, \ldots, C_{12}$ can be found by minimising the mean score $\overline{S(\tilde{\mathbf{p}}, \mathbf{o})}$ of the recalibrated forecasts.


{}

\label{lastpage}
\end{document}